\documentclass{aa}
\usepackage{graphicx}
\usepackage{amsmath}	
\usepackage{amssymb}	
\usepackage{relsize}
\usepackage{txfonts}
\usepackage{placeins}
\usepackage{natbib}
\usepackage[colorlinks=true,linkcolor=blue,allcolors=blue]{hyperref}%
\usepackage[english]{cleveref}
\crefname{figure}{Fig.}{Figs.}
\Crefname{figure}{Fig.}{Figs.}
\usepackage{rotating}
\usepackage{booktabs}
\usepackage{balance}
\usepackage{needspace}
\usepackage{xcolor}
\usepackage{multirow}
\usepackage{makecell} 
\usepackage{threeparttable}
\usepackage{tabularx}
\usepackage{dblfloatfix}

\newcommand{\thinhline}{\noalign{\hrule height 0.3pt}}

\newcommand{\Msun}{M$_\odot$}

\begin{document} 

   \title{Dynamically defined bulges reveal that early-type galaxies with compact bulges host more massive black holes}
   \subtitle{}

   \author{Hanjing Shi$^{1,2}$, Ling Zhu$^1$\thanks{lzhu@shao.ac.cn}, Sabine Thater$^3$, Yong Shi$^4$, Ana L. Chies-Santos$^5$, Luis Gabriel Dahmer Hahn$^6$, \\Zhengya Zheng$^1$}

   \institute{
$^{1}$ Shanghai Astronomical Observatory, Chinese Academy of Sciences, 80 Nandan Road, Shanghai 200030, China \\
$^{2}$ School of Astronomy and Space Science, University of Chinese Academy of Sciences, No.19A Yuquan Road, Beĳing 100049, China \\
$^{3}$ Department of Astrophysics, University of Vienna, Türkenschanzstraße 17, 1180 Wien, Austria\\
$^{4}$ Department of Astronomy, Westlake University, Hangzhou, Zhejiang 310030, China\\
$^{5}$Instituto de Física, Universidade Federal do Rio Grande do Sul, Av. Bento Gonçalves 9500, Porto Alegre, R.S. 90040-060, Brazil\\
$^{6}$Institute of Cosmology and Gravitation, University of Portsmouth, 1-8 Burnaby Road, Portsmouth, PO1 3FX, UK\\
}

   \date{Received; accepted}
   
   \titlerunning{Dynamically defined bulge}
\authorrunning{Shi et al.}  
 
  \abstract
   {We revisit the $\ M_{\rm BH}-M_{*,\rm bulge}$ relation for early-type galaxies by introducing a uniform bulge defined by dynamical decomposition. From the set of nearby galaxies with accurately measured black hole masses, we select 30 early-type galaxies that have been observed with wide-field integral-field unit instruments (MUSE, ATLAS$^{\rm 3D}$, or PPAK). For each galaxy, we construct an orbit-superposition Schwarzschild model and obtain its stellar orbit distribution. In this framework, we define the bulge uniformly as comprising stellar orbits with circularity $\lambda_z < 0.7$ (or, in an alternative definition, $\lambda_z < 0.2$) and lying within $r < 3$ kpc. From these orbits, we reconstruct the spatial density distribution of the bulge and measure its mass ($M_{*, \rm bulge}$), size ($R_{\rm e, bulge}$) and surface density slope ($\gamma_{\rm bulge}$). We find that the dynamical bulge defined by $\lambda_z < 0.7$ yields masses that are systematically consistent with the photometric bulge masses. Comparing to the bulge from photometric decomposition, using this dynamical bulge definition slightly decreases the scatter around the best-fitting from 0.56 to 0.47. More interestingly, we identify a clear trend: at a given bulge mass, galaxies hosting more compact bulges (i.e. smaller bulge sizes or steeper bulge density slopes) systematically contain more massive black holes; eight outliers show overmassive BHs together with ultra-compact bulges. We interpret this as direct observational evidence that the $M_{\rm BH}-M_{*,\rm bulge}$ relation evolves over a galaxy’s merger history, with that history imprinted in bulge compactness. Incorporating compactness via an $M_{\rm BH}-M_{*, \rm bulge}-R_{\rm e, bulge}$ (or $\gamma_{\rm bulge})$ relation brings the outliers into agreement and reduces the overall scatter of the sample to $\sim 0.3$.
   }

   \maketitle

\section{Introduction}
\label{sec:intro}

The tight correlations between supermassive black hole (SMBHs) masses and the properties of their host galaxy bulges — most notably the $M_\mathrm{BH}$--$\sigma$ relation \citep{Ferrarese_2000, Gebhardt_2000} and the $M_\mathrm{BH}$--$M_\mathrm{*,bulge}$ relation \citep{Marconi_2003, Haring_2004} — suggest that SMBHs and bulges undergo a co-evolutionary process regulated by a common physical mechanism, possibly AGN feedback \citep{DiMatteo_2005, Kormendy_2013}. SMBH masses also correlate tightly with the number of globular clusters in their host galaxies ($M_{\rm BH} \propto N_{\rm GC}$; \citealt{Burkert_2010, Harris_2011, Harris_2014}), likely reflecting a shared dependence on host mass and assembly history.

A reliable description of this co-evolution, therefore, hinges on precise determinations of both the BH mass ($M_\mathrm{BH}$) and the bulge stellar mass ($M_\mathrm{*,bulge}$). Considerable effort has been devoted to obtaining accurate BH masses using the kinematics of stars \cite[e.g.,][]{Gebhardt_2003_stellar,van_den_Bosch_2010_stellar,Thater2022MNRAS.509.5416T, Thater2023A&A...675A..18T, Nguyen2026ApJ..1003...98N}, ionised gas \cite[e.g.,][]{Macchetto_1997_ionized, Barth_2001_ionized,Walsh_2013_ionized}, and central CO discs \cite[e.g.,][]{Davis_2013_NA_CO, Barth_2016a_CO, Nguyen2020ApJ...892...68N, Ruffa_2024_CO}. These methods typically deliver BH mass measurements with uncertainties of about 0.1–0.3 dex. Results based on stellar kinematics and CO are generally in good agreement, although a systematic offset of $\sim$0.2 dex has been reported \citep{Thater2020IAUS..353..199T}. On the other hand, the bulge mass is usually derived through photometric bulge–disc decompositions \citep{Peng_2010,Erwin_2015}. It has been recognised for years that the bulge mass also carries substantial uncertainties; measurements from different groups can differ by roughly 0.3 dex in their bulge mass \citep{Saglia_2016}.

Although there is broad agreement that a correlation exists, the exact \textit{functional form} of the $M_\mathrm{BH}$--$M_\mathrm{*,bulge}$ relation, and whether it undergoes intrinsic evolution with galaxy morphology, remains highly contested. 
Previous studies have suggested that S\'{e}rsic and core-S\'{e}rsic spheroids occupy distinct sequences \citep{Graham_2012}, and that late- and early-type galaxies likewise follow different relations \citep{Savorgnan_redseq_2016,Kayhan2009ApJ...698..198G}. 
More recently, it has been shown that even within the early-type class, pure ellipticals (E) and lenticulars (ES/S0) obey \textit{distinct} $M_\mathrm{BH}$--$M_\mathrm{*,bulge}$ relations that share similar slopes \citep{Sahu_2019}. 
The traditionally adopted, nearly linear slope—reported to range from $0.85$ \citep{Saglia_2016} to $1.16$ \citep{Kormendy_2013}—is now recognised as largely arising from the heterogeneous morphological mix of the samples, rather than from a single universal physical relation \citep{Graham_2022}. 
Galaxy merger histories have been invoked as the mechanism driving the transformation of this relation from late- to early-type systems. 

Relic galaxies are rare, massive, compact systems that formed most of their stars rapidly at early epochs and experienced little subsequent stellar mass growth \citep{QuilisTrujillo_2013,FerreMateu_2017}. Their old stellar populations and compact structures resemble those of massive quiescent galaxies at $z\sim2$, often referred to as ``red nuggets'' \citep{vanDokkum_2008,Trujillo_2014_relic}. Typical massive galaxies continue to grow through the accretion of stars from other galaxies \citep{Oser_2010,vanDokkum_2010}. This makes Relic galaxies a useful population for examining how assembly history relates to the $M_{\rm BH}$--$M_{\rm bulge}$ relation. 
There is evidence that the ultracompact early-type galaxies (ETGs) in the local universe have overmassive BHs \citep{Walsh_2016_1277, van_den_Bosch_2016}.
However, the relic population is itself heterogeneous. NGC~0384, a nearby compact galaxy interpreted as a surviving relic galaxy, lies within the intrinsic scatter of the local $M_{\rm BH}$--$M_{\rm bulge}$ relation, whereas several other compact relics are significant positive outliers \citep{Cohn_2024_NGC384}. It was argued that compactness or relic status alone does not uniquely determine a galaxy's offset from the scaling relation; instead, the offset may reflect the balance between early BH growth and subsequent stellar assembly \citep{Cohn_2025}.


Crucially, however, how the bulge is defined strongly affects any inferred morphological evolution. When the bulge in elliptical galaxies is taken to be only the central component, with the outer envelope attributed to dissipationless mergers, fast and slow rotators were found to follow the same correlation between bulge mass and BH mass \citep{Zhu&Ho2021ApJ...907....6Z}. 
Given the limited information from photometric imaging alone, differing prior assumptions about the decomposition into bulge and disc can also yield contradictory interpretations as to whether ultra-compact ETGs host overmassive black holes and are thus outliers \citep{Walsh_2015_1271, Walsh_2016_1277, Walsh_2017_1216h, Cohn_2024_NGC384}, or instead are consistent with the standard relation \citep{Savorgnan_2016}.

The kinematic data offer complementary insights into the internal structure of galaxies. An early example was the SAURON survey, which mapped the stellar kinematics of 48 elliptical and lenticular galaxies and revealed features such as kinematic twists, decoupled components, and central discs \citep{Emsellem_2004, Emsellem_2007}. Building on this work, the ATLAS$^{\rm 3D}$ survey extended the study to a complete, volume-limited sample of 260 nearby ETGs, enabling a systematic census of their properties \citep{Cappellari_2011_Atlas3D_summarize, Cappellari_2016_review}.

Subsequent IFU surveys, including CALIFA \citep{CALIFA_2012}, MaNGA \citep{SDSS_2015}, and SAMI \citep{SAMI_2012}, have delivered kinematic maps for thousands of nearby galaxies. By jointly analysing photometric images and stellar kinematic maps from these IFU observations, dynamical modelling techniques—particularly orbit-superposition methods—have been developed to recover the internal distribution of stellar orbits in galaxies \citep{Schwarzschild_1979_model, Zhu_2018c_model, Vasiliev_2020, Zhang_2026_SchwarMAX}, enabling a decomposition of different structures based on the dynamical properties of the orbits, e.g., circularity and time-averaged radius. This approach has been applied to several hundred galaxies from CALIFA \citep{Zhu_2018a, Zhu_2018b}, MaNGA \citep{Jin_2021_MaNGA}, and SAMI \citep{Santucci_2022a,Santucci_2022b}. The method has subsequently been extended to simultaneously fit stellar kinematics and stellar population properties \citep{Zhu_2020, zhu_2022_NGC1380, Poci_2019_MGE3115}, achieving a chemo-dynamical decomposition of galactic structures (see review \citealt{vdV2025ARA&A..63..259V}). Such dynamical decompositions can effectively disentangle complex components, such as the classical bulge, bar, nuclear disc, and extended disc, and derive their distinct chemical properties, thereby clearly separating structures with different physical origins \citep{Tahmase2022ApJ...941..109T, Jin_2025_barredmodel, Jin2026arXiv260700497J}.

Among the galaxies with measured BH masses, a substantial subset has wide-field IFS data from the ATLAS$^{3D}$ survey, observed with the SAURON IFU ($33''\times41''$; \citealt{Cappellari_2011_Atlas3D_summarize}), from the PMAS/PPAK observations presented by \citet{Y_ld_r_m_2017}, or from MUSE in wide-field mode ($60''\times60''$; \citealt{Bacon_2010_VLT_MUSE}). PPAK is a wide-field fibre-bundle IFU mounted on the PMAS spectrograph at the Calar Alto 3.5-m telescope, with a hexagonal science field of approximately $74''\times64''$ \citep{Kelz_2006_PPAK}. These observations typically provide spatial coverage that encompasses the bulge-dominated regions.

In this work, we select a sample of 30 ETGs and construct a Schwarzschild orbit-superposition model for each one. We introduce a uniform dynamical definition of the bulge across the entire sample and measure its mass, size, and density slope. We demonstrate that for ETGs, the BH–bulge mass relation becomes tighter when this dynamically defined bulge is used, and notably, we find a clear evolutionary trend with the bulge compactness.
The paper is organised as follows. Section~\ref{sec:data} details the sample selection and the observational data. Section~\ref{sec:model} describes the modelling methodology and the orbital decomposition. The results are presented in Section~\ref{sec:analysis} and discussed in Section~\ref{sec:discussion}. Finally, Section~\ref{sec:conclusion} summarises our main conclusions.

\section{Data preparation}
\label{sec:data}
\subsection{Sample}
We focus on galaxies that have a directly measured black hole mass from spatially resolved dynamical modelling (either stellar or gas dynamics, or maser observations) in the literature, so that we can place each system on the $M_{\rm BH}$--$M_{\rm bulge}$ scaling relation without relying on indirect calibrations. 

To construct a well-controlled sample for dynamical bulge decomposition, we require each galaxy to have integral-field spectroscopic (IFS) observations extending beyond 2\,kpc from the galaxy centre. 

This coverage allows us to map the internal kinematic structure, enabling a robust separation of the bulge and disc components via the orbital circularity parameter $\lambda_z$; additionally, high-quality photometric imaging must be available to construct a Multi-Gaussian Expansion (MGE) model of the surface brightness distribution, which serves as the photometric input for the dynamical modelling. 

We further restrict our sample to early-type galaxies (ellipticals and lenticulars, ES/S0) with total stellar masses in the range $\log(M_*/M_\odot) \approx [10.3$--$11.6]$. This regime includes galaxies with both pressure-supported bulges and rotating discs, and their location on the $M_{\rm BH}$--$M_{\rm bulge}$ shows a large scatter with the photometric bulge \citep{Graham_2022}. We excluded the most massive giant elliptical galaxies with $M_*\gtrsim 10^{12}\,$\Msun, which have a cored surface brightness profile and a large spheroid size, of which only a small fraction is covered by the IFU data. All the galaxies included in our sample have relatively steep surface brightness profiles in the central regions; i.e., they are not obviously cored ellipticals.

Applying these criteria, we assemble a final sample of 30 ETGs drawn from three complementary datasets: the ATLAS \textsuperscript{3D} survey (\citeauthor{Cappellari_2011_Atlas3D_summarize} \citeyear{Cappellari_2011_Atlas3D_summarize}), archival MUSE observations, and PPAK observations. The basic properties of the galaxies are summarised in Table~\ref{tab:bulge_properties}.

\subsubsection{ATLAS\textsuperscript{3D}}
The ATLAS$^{\rm 3D}$ Project combines a multi-wavelength survey of a complete sample of 260 early-type galaxies within the local (42 Mpc) volume ($1.16 \times 10^{5} Mpc^{3}$) with numerical simulations and semi-analytic modelling of galaxy formation. This survey used SAURON integral-field observations, the data were Voronoi-binned (\citeauthor{Cappellari_2003_vorbin} \citeyear{Cappellari_2003_vorbin}) to achieve a target signal-to-noise ratio of S/N = 40 for all galaxies. We selected 20 galaxies that meet our selection criteria; detailed information is included in Table~\ref{tab:bulge_properties}.

\subsubsection{MUSE}
The integral-field spectroscopic data used in this work were obtained with the VLT/MUSE instrument \citep{Bacon_2010_VLT_MUSE} in Wide-Field Mode without adaptive optics (WFM-NOAO). For NGC 3115, the observations consist of a five-point mosaic covering the nominal wavelength range of ~480–930 nm, with a spatial sampling of 0.2 arcsec per pixel and a 1'× 1' field of view per pointing. The mosaic extends to a maximum galactocentric radius of ~120 arcsec (~5.6 kpc) along the major axis, corresponding to approximately 2.5 effective radii ($R_e$).  The reduced datacubes are publicly available through the CDS archive \citep{Gu_rou_2016_MUSE_ngc3115}.

For NGC~0584, we adopted the large-scale stellar kinematics derived from VLT/MUSE observations and presented by \citet{Thater_2019}. For NGC~1332, NGC~4526, and NGC~6861, we used the data reduced by \citet{Zhang_2026} using the standard pipeline described in \citet{Weilbacher_2020_MUSE_pipeline}.

\subsubsection{PPAK}
We include six compact early-type galaxies: MRK~1216, NGC~0384, NGC~1271, NGC~1277, PGC~11179, and UGC~2698, selected from the merger-free sample of \cite{Zhu_2025_merger_free}. These galaxies were originally studied by \citet{Y_ld_r_m_2017} using high-resolution \textit{HST} imaging together with wide-field integral-field spectroscopy obtained with the PPAK IFU, the wide-field fibre-bundle mode of the PMAS spectrograph on the Calar Alto 3.5-m telescope \citep{Verheijen_2004_PPAK,Kelz_2006_PPAK}. They are characterised by compact morphologies, old stellar populations, and high stellar velocity dispersions; therefore, they are included as the compact ETG subset of our sample.

\subsection{Stellar kinematics}
Although the IFU data were from different observations, the stellar kinematics adopted were all obtained using the
publicly available penalised pixel-fitting code \textsc{pPXF}
\citep{Cappellari_2004_ppxf,Cappellari_2016_ppxf}. \textsc{pPXF}
performs full-spectrum fitting: each observed spectrum is modelled as
a linear combination of stellar templates from the empirical MILES
or E-MILES libraries
\citep{Sanchez_Blazquez_2006_MILES,Vazdekis_2016_EMILES}, broadened
by a line-of-sight velocity distribution (LOSVD) that is expanded in
a Gauss--Hermite series. The expansion coefficients are the mean
line-of-sight velocity $V$, the velocity dispersion $\sigma$, and the
higher-order moments $h_3$ and $h_4$, which describe the asymmetric
and symmetric deviations of the LOSVD from a Gaussian. The observed
spectrum is fitted simultaneously across the wavelength window
adopted for each dataset. 

The details of the template library,
the fitted wavelength range, and the binning of each dataset may be slightly different and are given in the corresponding data papers cited above.

\begin{figure*}
     \centering
     \includegraphics[width=1\linewidth]{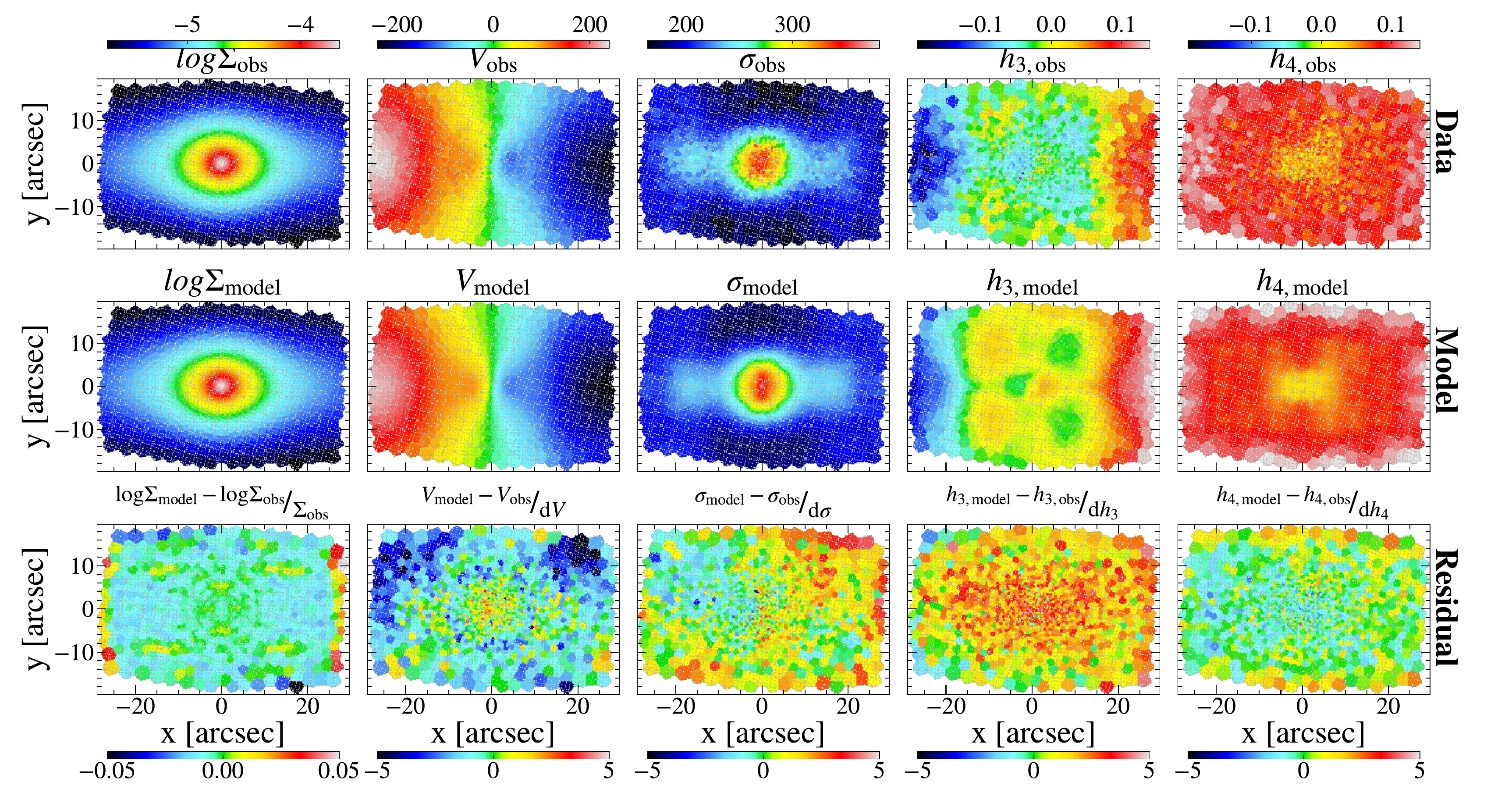}
     \caption{The best-fitting Schwarzschild model of the galaxy NGC~1332. Columns from left to right show the 2D surface brightness ($\Sigma$), velocity ($V$), velocity dispersion ($\sigma$), $h_3$, and $h_4$. The first row displays the observed surface brightness and stellar kinematics. The second row shows the corresponding best-fit Schwarzschild model. The third row the normalised residuals $(\mathrm{data} - \mathrm{model})/\mathrm{error}$, in units of the observational uncertainty.}
     \label{fig:kinematic_map}
 \end{figure*} 


\subsection{Multi-Gaussian Expansion (MGE) photometric models}
\label{sec:mge}
Accurate photometric models of the stellar surface brightness 
distribution are required as inputs to the dynamical modelling.
We parameterize the surface brightness of each galaxy using the 
Multi-Gaussian Expansion \citep[MGE;][]{Emsellem_1994_MGE} method, 
as implemented in the \textsc{MgeFit} package of 
\citet{Cappellari_2002_MGE}.
In this approach, the projected surface brightness is decomposed 
into a sum of two-dimensional Gaussians,

\begin{equation}
    I(x, y) = \sum_{j=1}^{N} I_j \exp\left(-\frac{1}{2\sigma_j^2}
    \left[x'^2 + \frac{y'^2}{q_j^{\prime\,2}}\right]\right),
    \label{eq:mge}
\end{equation}

\noindent where $x'$ and $y'$ are the coordinates along the major 
and minor axes of the galaxy, respectively; $I_j$, $\sigma_j$, and 
$q'_j$ are the peak surface brightness, standard deviation along 
the major axis, and projected axial ratio of the $j$-th Gaussian 
component; and $N$ is the total number of components.

We adopt MGE models from the literature for all our sample except NGC\,1332. For the 19 galaxies drawn from the ATLAS\textsuperscript{3D} 
survey, we use the MGE models of \citet{Scott_2013_ATLAS_MGE}, 
which were fitted to SDSS $r$-band or INT $r$-band images.
For the six PPAK galaxies (MRK\,1216, NGC\,0384, NGC\,1271, NGC\,1277, PGC\,11179,
and UGC\,2698), we use the MGE models presented in 
\citet{Zhu_2025_merger_free}, derived from HST imaging.
For the MUSE subsample, we use the models of 
\citet{Emsellem_1999_MGE3115} as adopted in 
\citet{Poci_2019_MGE3115} for NGC\,3115, 
\citet{Scott_2013_ATLAS_MGE} for NGC\,4526, and 
\citet{Kabasares_2022_MGE6861} for NGC\,6861.
The only exception is NGC\,1332, for which no suitable published 
MGE model was available, and we created the MGE ourselves (see Appendix). The resulting MGE parameters of NGC\,1332 are listed in Table~\ref{tab:mge_ngc1332}.
The image and MGE model for all 30 galaxies are included in \cref{fig:25image}.

\begin{figure*}
    \centering
    \includegraphics[width=1\linewidth]{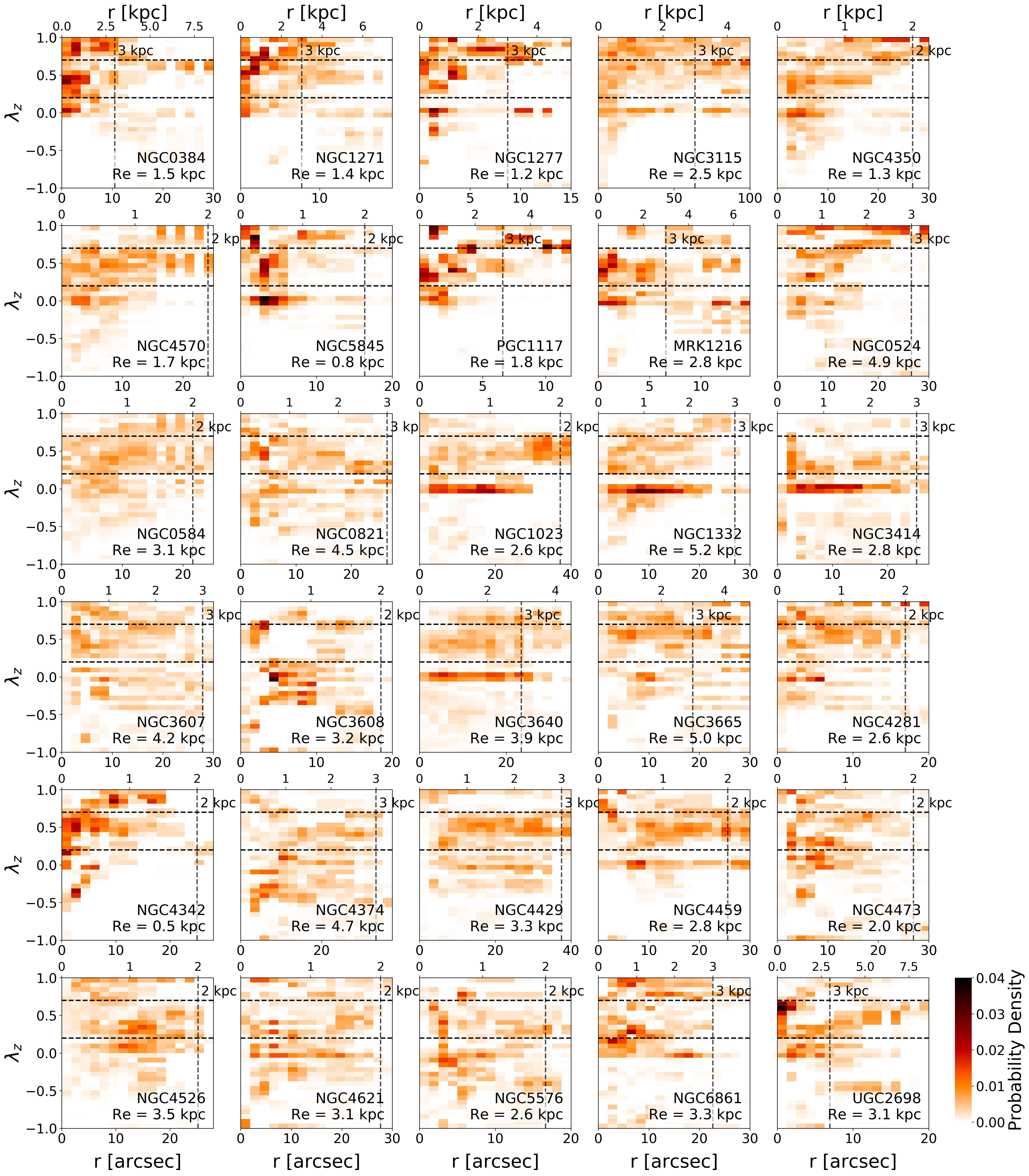}
    \caption{Stellar orbit distributions of the 30 galaxies in the sample, derived from the best-fitting Schwarzschild orbit-superposition models. Each panel shows the probability density of stellar orbits in the phase space of time-averaged radius $r$ versus circularity $\lambda_z$, with the colour scale as indicated by the colour bar. The galaxy name and effective radius $R_\mathrm{e}$ are labelled in each panel; $R_\mathrm{e}$ values are taken from \citet{Cappellari_2011_Atlas3D_summarize} for galaxies in the ATLAS$^\mathrm{3D}$ sample, and from \citet{Rusli_2011}, \citet{Emsellem_1999}, and \citet{Rusli_2013} for NGC\,1332, NGC\,3115, and NGC\,6861, respectively. The two horizontal dashed lines mark the circularity thresholds $\lambda_z = 0.2$ and $\lambda_z = 0.7$ corresponding to two definitions of bulge (see Sect.~\ref{sec:bulge_def}). The vertical dashed line indicates the adopted radial boundary $r_\mathrm{cut}$, which is set to  3 kpc for 2/3 of the sample and 2 kpc for the remaining 1/3. 
    }
    \label{fig:circularity25}
\end{figure*}

\section{Orbit-superposition model and dynamical structure decomposition}
\label{sec:model}
We model the stellar dynamics of 30 galaxies using the orbit-superposition Schwarzschild method \citep{Schwarzschild_1979_model}, as developed by \citet{vandenBosch_2008_model} and made public as the dynamite code \citep{Jethwa2020ascl.soft11007J, Thater2022A&A...667A..51T}. This approach constructs a library of stellar orbits within a given gravitational potential and determines the orbital weight distribution by simultaneously fitting the observed two-dimensional luminosity distributions and stellar kinematics  (see Section~\ref{sec:schwarzschild}). 

\subsection{Orbit superposition Schwarzschild Modelling}
\label{sec:schwarzschild}
The major steps in constructing the orbit superposition model include:(1) construction of a model for the gravitational potential with a few free parameters, (2) calculation of an orbit library in the gravitational potential, and (3) inferring the weights of orbits by fitting the model to the data, including the luminosity density and kinematic maps. We calculate hundreds of models by exploring the free parameters in the gravitational potential and find the best-fitting models with a minimum $\chi^2$ between the model and the data. In the end, we obtain 3D models superposed by stellar orbits that match all the data well.

\subsubsection{Gravitational potential}
\label{sec:freeparameter}
The gravitational potential is generated by a combination of the stellar, BH, and dark matter (DM) distributions.
To derive a model for the stellar mass, we first deprojected the surface brightness described by the MGE model (see Section~\ref{sec:mge}) to obtain the intrinsic luminosity density distribution by assuming a set of viewing angles $(\vartheta, \varphi, \psi)$ for the system and then multiplying by the stellar mass-to-light ratio $\Upsilon_* \equiv M_*/L$.

The intrinsic three-dimensional shape of the stellar mass 
distribution is parameterized by two axial ratios, 
$p \equiv b/a$ and $q \equiv c/a$, where $a \geq b \geq c$ 
denotes the lengths of the major, intermediate, and minor 
principal axes, respectively, and a scaling parameter $u = \sigma_{\rm obs}/\sigma_{\rm intr}$, which is the ratio of the long axis in the projected surface brightness to the long axis of the 3D intrinsic density distribution.

The intrinsic shape parameters $(p, q, u)$ are directly linked to the deprojection of the MGE surface brightness model by a set of viewing angles. Given the projected axial ratios $q'_j$ and the photometric major axis $\sigma'_j$ of the MGE, a choice of $(p, q, u)$ uniquely determines the three viewing angles $(\vartheta, \varphi, \psi)$ -- the inclination, azimuthal angle, and roll angle, respectively -- via the deprojection equations of \citet{vandenBosch_2008_model}. The MGE model is then analytically deprojected along these viewing angles to yield the three-dimensional luminosity density, which is subsequently converted to a stellar mass density using the mass-to-light ratio $\Upsilon_*$.

All three parameters ($p$, $q$, $u$) and the stellar mass-to-light ratio $M/L_*$ are treated as free parameters within physically motivated ranges and are simultaneously constrained by fitting the observed two-dimensional kinematic maps within the Schwarzschild modelling framework. The allowed ranges and adopted sampling strategies for each parameter are described in Section~\ref{sec:freeparameter}.

The DM component is modelled as a spherical NFW halo \citep{Navarro_1997_NFW}, with the concentration parameter $c$ fixed following the halo mass–concentration relation of \citet{Dutton_2014_c}:
\begin{equation}
    \log_{10} c = 0.905 - 0.101\log_{10}
    \left(\frac{M_{200}}{10^{12}\,h^{-1}\,M_{\odot}}\right),
\end{equation}
with $h = 0.671$ \citep{Planck_Collaboration_2014}. Fixing 
$c$ in this way does not significantly affect the enclosed 
DM profile within the spatial extent probed by the PPAK/ATLAS\textsuperscript{3D}/ MUSE data and, therefore, does not introduce a systematic bias in the inferred $M_*/L$.

The black hole mass $M_{\rm BH}$ is fixed at the value measured from the literature for each galaxy individually (see Table~\ref{tab:bulge_properties}), as the spatial resolution of the IFU data is insufficient to resolve the black hole sphere of influence. Consequently, the data do not constrain $M_{\rm BH}$ through its effect on the stellar kinematics at the spatial scales probed here. And fixing it to the literature value does not bias the inferred $M_*/L$ or the orbital structure.

In total, we have five free hyper-parameters in the gravitational potential: the stellar mass-to-light ratio $M_*/L$, the DM halo mass $M_{200}$, and the three intrinsic shape parameters $q$, $p$, $u$.

\subsubsection{The orbit library}
In each given potential, the orbit library is sampled on a grid of $n_E \times n_{\theta} \times n_R$ initial conditions, where $n_E$ is the number of logarithmically spaced energy shells, $n_{\theta}$ is the number of angular momentum orientations, and $n_R$ is the number of orbital shapes sampled in the meridional plane. The sampling is chosen to match the spatial resolution and the number of kinematic constraints provided by each dataset. For the ATLAS$^{\rm 3D}$ galaxies, we adopt $n_E \times n_{\theta} \times n_R = 21 \times 11 \times 7$; for the MUSE galaxies, $21 \times 15 \times 11$; and for the PPAK galaxies, $15 \times 9 \times 7 $, following \citet{Zhu_2018c_model}. 

Each set of initial conditions is dithered by a factor of 3, giving a total orbit library of $n_E \times n_{\theta} \times n_R \times \mathrm{dithering}^3$ orbits per model. This sampling has been shown to provide a converged solution for datasets of comparable resolution and kinematic constraint density \citep{vandenBosch_2008_model, Zhu_2018c_model}.

\balance
\subsubsection{Best-fitting Model}
For each assumed gravitational potential, we determine the orbital weights by minimizing the $\chi^2$ between the model and the data. The quality of the fit is then assessed by comparing the model predictions with the observed two-dimensional stellar kinematics, namely the mean velocity $V$, velocity dispersion $\sigma$, and the higher-order Gauss-Hermite moments $h_3$ and $h_4$.

We determine the best-fitting model by exploring the parameter space of the dynamical models and identifying the minimum $\chi^2$.
The exploration is performed iteratively. We begin with an initial grid of models spanning five free parameters mentioned in Section~\ref{sec:freeparameter} ($M_*/L$, $M_{200}$, 
$q$, $p$, $u$). Models with relatively low $\chi^2$ values are then selected, and additional models are constructed in their vicinity by stepping through the parameter space. This procedure is repeated, progressively refining the sampling around the best-fitting regions, until the minimum $\chi^2$ is well constrained and all surrounding models have been sufficiently explored.

The model with the minimum $\chi^2$ ($\chi^2_{\rm min}$) is adopted as the fiducial model. This best-fitting model is used to derive the orbital decomposition and to compute the bulge stellar mass, as described in Section~\ref{sec:bulge_def}.

Confidence intervals are defined based on the $\Delta\chi^2$ relative to this minimum value. The $1\sigma$ confidence level is defined as
\begin{equation}
\Delta\chi^2 \equiv \chi^2 - \chi^2_{\rm min} <
\sqrt{2 \times n_{\rm GH} \times N_{\rm obs}},
\end{equation}
where $n_{\rm GH} = 4$ is the number of kinematic moments ($V$, $\sigma$, $h_3$, $h_4$), and $N_{\rm obs}$ is the number of spatial bins in the kinematic maps.

Models within the $1\sigma$ confidence level are used to estimate the uncertainties on the derived quantities, in particular, the bulge stellar mass. The exploration of the parameter space is illustrated in \cref{fig:freepara} for NGC~1332 as a representative example, and the corresponding best-fitting kinematic maps are shown in \cref{fig:kinematic_map}.

\subsubsection{Stellar orbit distribution}
We characterise each orbit by two properties: the time-averaged radius $r$, which represents the spatial extent of the orbit, and the circularity $\lambda_z \equiv J_z / J_{\rm max}(E)$, where $J_z$ is the orbit's angular momentum about the galaxy's short axis and $J_{\rm max}(E)$ is the maximum angular momentum of a circular orbit with the same specific orbital energy, $E = \frac{1}{2}|\mathbf{v}|^2 + \Phi(\mathbf{x})$. Here, $\mathbf{v}$ and $\Phi(\mathbf{x})$ denote the orbit's velocity and the gravitational potential at position $\mathbf{x}$, respectively.
Dynamically cold, rotationally supported orbits have $|\lambda_z| \sim 1$, while dynamically hot orbits dominated by random motions have $\lambda_z \sim 0$. Orbits with $\lambda_z < 0$ counter-rotate with respect to the net prograde rotation of the galaxy.

We describe the stellar orbit distribution of a model as a probability density $p(r, \lambda_z)$ in the phase space of time-averaged radius $r$ versus circularity $\lambda_z$, following \citep{Zhu_2018c_model}. The stellar orbit distribution of the best-fitting model for all 30 galaxies is shown in \cref{fig:circularity25}.

\balance
\subsection{Dynamical structure decomposition}
\label{sec:bulge_def}
We decompose each galaxy into three structures using thresholds in circularity, $\lambda_{z,\mathrm{cut}}$, and time-averaged radius, $r_{\mathrm{cut}}$:
\begin{itemize}
\item Disc : $\lambda_z>\lambda_{z,\mathrm{cut}}$
\item Bulge: $\lambda_z<\lambda_{z,\mathrm{cut}}$ and $r<r_{\rm cut}$
\item Halo: $\lambda_z<\lambda_{z,\mathrm{cut}}$ and $r>r_{\rm cut}$.
\end{itemize}
 We adopt two different definitions of bulge depending on the choice of $\lambda_{z,\mathrm{cut}}$, as illustrated in Fig.~\ref{fig:circularity25}. When we set $\lambda_{z,\mathrm{cut}} = 0.7$, we remove only the dynamically cold disc, while the bulge includes all dynamically warm ($0.2 < \lambda_{z} < 0.7$) and hot ($\lambda_{z} < 0.2$) orbits. 
 In contrast, when we adopt $\lambda_{z,\mathrm{cut}} = 0.2$, the bulge is limited to the dynamically hot, pressure-supported component, yielding the most conservative estimate of $M_\mathrm{bulge}$.

We further distinguish between the bulge and the halo by adopting a cutoff radius of $r_{\rm cut} = 3$ kpc, guided by both theoretical arguments and observations. Hydrodynamical simulations from TNG50 \citep{Pillepich_2019} show that dynamically hot components can arise from different processes: stars in the compact inner bulge are predominantly formed in situ, whereas stars in the more extended, hot inner stellar halo are mainly assembled through mergers. In most TNG50 galaxies, the transition between the compact inner bulge and the extended hot inner stellar halo occurs at $\sim3.5$ kpc for most galaxies with $M_*\gtrsim 10^{10.6}$\Msun\, and at $\sim 2$ kpc in galaxies less massive \citep{zhu_2022_NGC1380, Zhu2022A&A...660A..20Z}. We therefore define the bulge as the region within $r_{\rm cut} = 3$ kpc by default; however, there are a few galaxies with data coverage $<3$ kpc, and their orbital distribution beyond the data coverage is not well constrained; thus we chose $r_{\rm cut} = 2$ kpc for those galaxies. We will discuss the implications of this choice in the discussion section. Therefore, we aim to primarily include the in-situ bulge mass.

\balance
\subsection{The mass of the bulge and its uncertainties}

\begin{figure*}
    \centering
    \includegraphics[width=14cm]{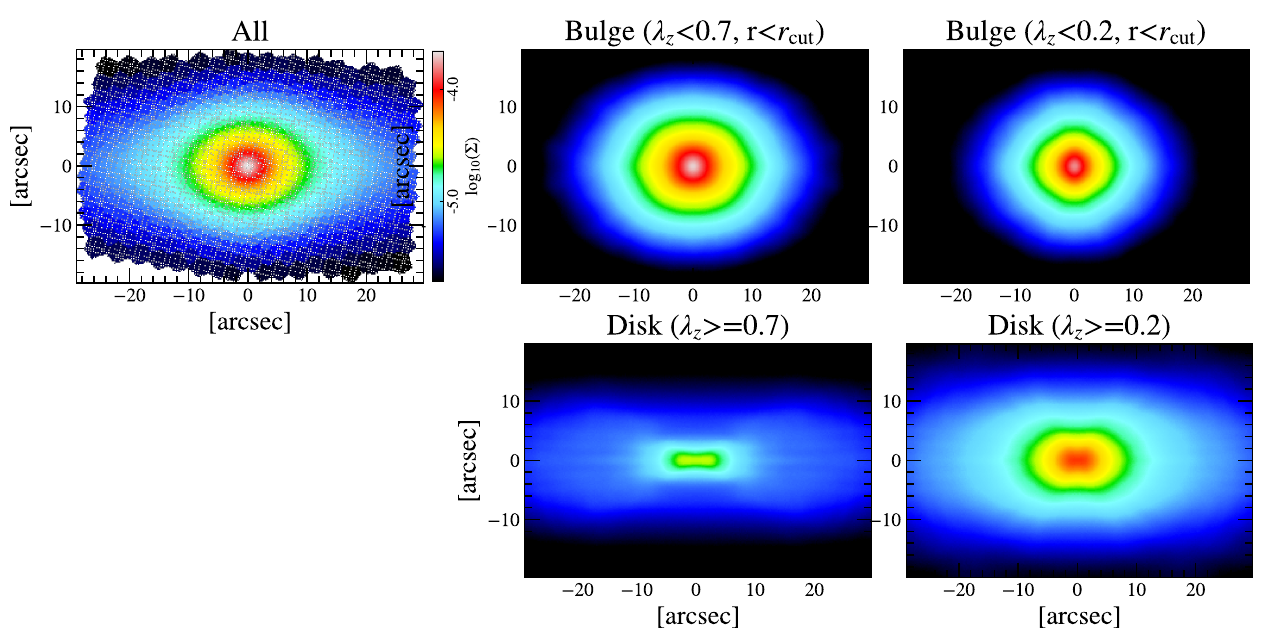}
    \caption{Normalised projected stellar surface-density maps of NGC~1332 reconstructed from the Schwarzschild orbit-superposition model. The top-left panel shows the total galaxy. The middle and right columns show the bulge and disc components obtained using the $\lambda_z<0.7$ and $\lambda_z<0.2$ dynamical definitions, respectively. Each map is normalised by the total projected luminosity of the corresponding component. All maps are projected onto the sky plane with the same orientation as the observational data. }
    \label{fig:2D_decomposition}
\end{figure*}

Once the disc, bulge, and halo are separated in $\lambda_z$ and $r$, we can use their orbits to reconstruct the density distribution of each component. \cref{fig:2D_decomposition} shows the surface brightness distribution of the disc and bulge of NGC~1332 projected onto the observational plane for two different choices of the separation in $\lambda_z$ for defining the bulge and disc.

Notice that the IFU data usually have limited spatial coverage, so the combination of the disc, bulge, and halo defined here does not necessarily equal the total luminosity of the galaxy. To set the absolute surface brightness of each component, we compute a scaling factor $w$ by matching the total model surface-brightness profiles to the MGE-derived profiles along both the major and minor axes. We then use this factor to set the absolute surface-brightness scale of each model component. The total luminosity of each component can thus be obtained. One example of the resulting scaled surface brightness profiles along the major and minor axes is illustrated in \cref{fig:bulge_decompositon}.

The bulge mass is then calculated by multiplying the bulge luminosity by the stellar mass-to-light ratio:
\begin{equation}
    M_{\rm bulge} = L_{\rm bulge} \times 
    \left(\frac{M_*}{L}\right)_{\rm dyn},
\end{equation}
where $L_{\rm bulge}$ denotes the total luminosity of bulge, and $\left(\frac{M_*}{L}\right)_{\rm dyn}$ is stellar mass-to-light ratio obtained by our best-fitting models.
This method is applied separately for the two definitions of bulge, taking $\lambda_z < 0.2$ or $\lambda_z < 0.7$, providing two independent estimates of $M_{\rm bulge}$ for each galaxy. The consequences of this choice for the inferred bulge masses will be examined in Section~\ref{sec:analysis}.

We assumed a spatially constant stellar mass-to-light ratio $M_*/L$ 
throughout each galaxy. Although stellar population gradients are 
known to exist in galaxies, with associated radial variations in 
$M_*/L$ \citep[e.g.,][]{Peletier_1990_agegradient,
Parikh_2021_agegradient}, the radial variation might be relatively small for ETGs with uniformly old stellar populations. In addition, if strong radial variation indeed exists in the stellar mass-to-light ratio, the dynamical model assuming constant $M_*/L$ will converge to the stellar $M_*/L$ in the very inner regions and bias the outer regions with the compensation of DM mass \citep{Lei2026arXiv260916740L}. The stellar $M_*/L$ of the bulge will thus be less affected.

We estimate the uncertainty in $M_{\rm bulge}$ directly from the 
ensemble of Schwarzschild models within the adopted $1\sigma$ 
confidence region around the best-fitting solution. For each accepted 
model, we calculate the bulge mass using the bulge luminosity and 
stellar mass-to-light ratio derived from that same model. This approach 
self-consistently propagates variations in the orbital weights, the 
resulting bulge luminosity fraction, and the stellar mass-to-light 
ratio while preserving their covariance. We adopt the minimum and 
maximum bulge masses among the accepted models as the lower and upper 
confidence limits, respectively, and report them as asymmetric 
uncertainties relative to the bulge mass of the best-fitting model.
The bulge mass estimated for each galaxy is listed in Table~\ref{tab:bulge_properties}. 

\begin{figure*}
    \centering
    \includegraphics[width=16cm]{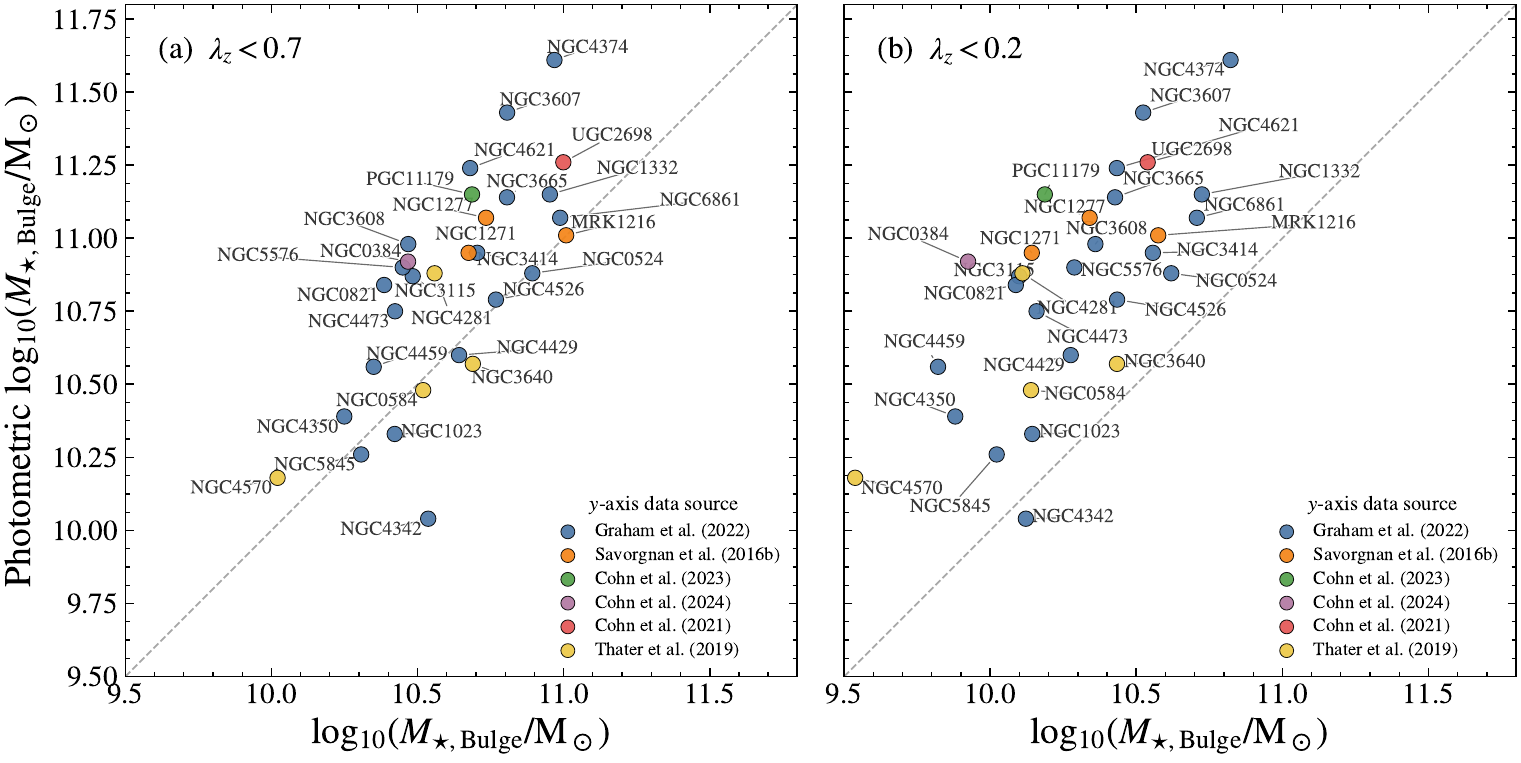}
    \caption{Comparison of dynamical and photometric bulge stellar 
    masses for the 30 sample galaxies. The horizontal axis shows 
    the dynamical bulge mass $M_{\rm *,bulge}$ derived in this 
    work; the vertical axis shows the photometric bulge masses compiled from \citep{Graham_2022}, \citep{Savorgnan_2016_bulgemass}, and the supplementary sources described in Section~\ref{sec:scaling}. Panels~(a) and (b) correspond to the $\lambda_z < 0.7$ and $\lambda_z < 0.2$ bulge definitions, respectively. The grey dashed line indicates the 1:1 relation. 
    }
    \label{fig:bulge_mass_compare2panles}
\end{figure*}

\begin{sidewaystable*}[htbp]
\setlength{\tabcolsep}{4.0pt}
\centering
\begin{threeparttable}
\caption{
Basic properties of the galaxy sample and bulge structural parameters derived from dynamical decomposition under two bulge selection criteria.
}
\label{tab:bulge_properties}
\begin{tabular}{l c c c c c c c | c c c | c c c }
\hline\hline
\noalign{\smallskip}
& & & & & & & &
\multicolumn{6}{c}{\textbf{Bulge}} \\[2pt]
\cmidrule(lr){9-14}
Galaxy &
IFU Data &
$D$ &
$\log M_{*,\rm dyn}$ &
$R_e$ &
$\log(M_{\rm BH}/M_\odot)$ &
$M_{\rm BH} Ref.$ &
$r_{\rm cut}$ &
\multicolumn{3}{c}{$\lambda_z < 0.7$} &
\multicolumn{3}{c}{$\lambda_z < 0.2$} \\
\cmidrule(lr){9-11} \cmidrule(lr){12-14}
& & & & & & & &
$R_{\rm e,bulge}$ & $\gamma_{\rm bulge}$ & $\log (M_{\rm bulge}/M_\odot)$ &
$R_{\rm e,bulge}$ & $\gamma_{\rm bulge}$ & $\log (M_{\rm bulge}/M_\odot)$ \\
& &  [Mpc] & & [kpc]& & & [kpc] & [kpc] & & & [kpc] & & \\
\hline
(1) & (2) & (3) & (4) & (5) & (6) & (7) & (8) & (9) & (10) & (11) & (12) & (13) & (14)\\
\hline
\noalign{\smallskip}
MRK~1216 & PPAK  
& 98.0 & 11.44 & 2.8 & $9.69\pm0.15$ &\citet{Walsh_2017_1216h} & 3.0
& 0.92 & 1.43 & 11.01
& 0.86 & 1.23 & 10.57 \\
NGC~0384 & PPAK
& 55.0 & 10.90 & 1.5 & $8.86\pm0.10$ &\citet{Cohn_2024_NGC384} & 3.0
& 0.69 & 1.73 & 10.47
& 0.69 & 1.54 & 9.92 \\
NGC~0524 & ATLAS\textsuperscript{3D} 
& 27.7 & 11.42 & 4.9 & $9.00\pm0.10$ &\citet{Krajnovi_NGC0524} &3.0
& 1.03 & 1.06 & 10.89 
& 1.15 & 1.12 & 10.62 \\
NGC~0584 & MUSE 
& 19.1 & 10.82 & 3.06 & $8.13\pm0.20$ &\citet{Thater_2019} &2.0
& 0.72 & 1.51 & 10.52 
& 0.55 & 1.83 & 10.14 \\
NGC~0821 & ATLAS\textsuperscript{3D} 
& 23.2 & 10.81 & 3.5 & $7.59 \pm 0.17$ &\citet{Gebhardt_2003_stellar}  & 3.0 
& 0.74 & 1.30 & 10.39 
& 0.65 & 1.36 & 10.09 \\
NGC~1023 & ATLAS\textsuperscript{3D} 
& 11.0 & 10.67 & 2.6 & $7.62\pm0.05$ &\citet{Bower_NGC1023} &2.0
& 0.61 & 1.65 & 10.42
& 0.47 & 1.78 & 10.14 \\
NGC~1271 &  PPAK 
& 85.8 & 11.11 & 1.4 & $9.48\pm0.16$ &\citet{Walsh_2015_1271} &3.0
& 0.78 & 1.65 & 10.67
& 0.83 & 1.44 & 10.14 \\
NGC~1277 &  PPAK 
& 73.0 & 11.17 & 1.2 & $9.69\pm0.14$ &\citet{Walsh_2016_1277} &3.0
& 0.71 & 1.64 & 10.73
& 0.62 & 1.69 & 10.34 \\
NGC~1332 & MUSE 
& 22.0 & 11.28 & 5.2 & $9.15\pm0.07$ &\citet{Rusli_NGC1332} &3.0
& 0.71 & 1.51 & 10.95
& 0.71 & 1.49 & 10.72 \\
NGC~3115 & MUSE 
& 9.3 & 10.89 & 2.5 & $8.94\pm0.25$ &\citet{Emsellem_NGC3115}&3.0
& 0.57 & 1.50 & 10.48
& 0.39 & 1.76 & 10.10 \\
NGC~3414 & ATLAS\textsuperscript{3D} 
& 24.3 & 11.08 & 3.0 & $8.38 \pm 0.09$ &\citet{Cappellari2008}&3.0 
& 0.84 & 1.41 & 10.70 
& 0.84 & 1.37 & 10.56 \\
NGC~3607 & ATLAS\textsuperscript{3D} 
& 25.0 & 11.23 & 7.9 & $8.16 \pm 0.18$ &\citet{Gultekin_2009}& 3.0 
& 1.20 & 1.11 & 10.81
& 1.25 & 1.03 & 10.52 \\
NGC~3608 & ATLAS\textsuperscript{3D} 
& 22.1 & 10.80 & 4.6 & $8.30 \pm 0.17$ &\citet{Gebhardt_2003_stellar}& 2.0 
& 0.74 & 1.44 & 10.47
& 0.75 & 1.45 & 10.36 \\
NGC~3640 & ATLAS\textsuperscript{3D} 
& 26.3 & 11.07 & 4.9 & $7.89 \pm 0.47$ &\citet{Thater_2019}& 3.0 
& 1.16 & 1.21 & 10.69
& 1.01 & 1.29 & 10.43 \\
NGC~3665 & ATLAS\textsuperscript{3D}  
& 34.7 & 11.37 & 5.0 & $8.76\pm0.10$ &\citet{Onishi_NGC3665} &3.0
& 1.23 & 0.92 & 10.81
& 1.18 & 0.96 & 10.43 \\
NGC~4281 & ATLAS\textsuperscript{3D}  
& 24.4 & 10.96 & 2.9 & $8.73\pm0.07$ &\citet{Thater_2019} &2.0
& 0.78 & 1.58 & 10.56
& 0.48 & 2.02 & 10.11 \\
NGC~4342 & ATLAS\textsuperscript{3D}  
& 22.8 & 10.73 & 0.5 & $8.65\pm0.18$ &\citet{Cretton1999} &2.0
& 0.39 & 2.33 & 10.54
& 0.29 & 2.42 & 10.12 \\
NGC~4350 & ATLAS\textsuperscript{3D}  
& 17.0 & 10.63 & 1.3 & $8.87\pm0.41$ &\citet{Pignatelli_NGC4350} &2.0
& 0.48 & 1.82 & 10.25
& 0.40 & 1.87 & 9.88 \\
NGC~4374 & ATLAS\textsuperscript{3D} 
& 17.7 & 11.35 & 11.0 & $8.95 \pm 0.05$ &\citet{Walsh_NGC4374} & 3.0 
& 0.79 & 1.19 & 10.97 
& 0.73 & 1.25 & 10.82 \\
NGC~4429 & ATLAS\textsuperscript{3D}  
& 16.6 & 10.93 & 3.3 & $8.18\pm0.08$ &\citet{Davis_NGC4429} &3.0
& 1.25 & 0.97 & 10.64
& 1.07 & 1.21 & 10.28 \\
NGC~4459 & ATLAS\textsuperscript{3D} 
& 15.5 & 10.63 & 2.8 & $7.82\pm0.10$ &\citet{Sarzi_NGC4459} &2.0
& 0.93 & 1.32 & 10.35
& 0.69 & 1.45 & 9.82 \\
NGC~4473 & ATLAS\textsuperscript{3D} 
& 15.1 & 10.68 & 2.7 & $8.07 \pm 0.36$ &\citet{Gebhardt_2003_stellar} & 2.0 
& 0.70 & 1.49 & 10.42 
& 0.76 & 1.45 & 10.16 \\
NGC~4526 & MUSE 
& 16.3 & 11.14 & 3.5 & $8.65\pm0.04$ &\citet{Davis_NGC4526} &2.0
& 0.97 & 1.13 & 10.77
& 0.87 & 1.16 & 10.43 \\
NGC~4570 & ATLAS\textsuperscript{3D}  
& 17.1 & 10.35 & 1.49 & $7.83\pm0.15$ &\citet{Thater_2019} &2.0
& 0.56 & 1.79 & 10.02
& 0.32 & 2.77 & 9.54 \\
NGC~4621 & ATLAS\textsuperscript{3D} 
& 17.6 & 10.99 & 7.6 & $8.59 \pm 0.06$ &\citet{Cappellari2008} & 2.0 
& 1.06 & 1.33 & 10.68 
& 0.84 & 1.45 & 10.43 \\
NGC~5576 & ATLAS\textsuperscript{3D}
& 24.5 & 10.74 & 5.8 & $8.19 \pm 0.10$ &\citet{Gultekin_2009} & 2.0 
& 0.81 & 1.49 & 10.45 
& 0.84 & 1.56 & 10.29 \\
NGC~5845 & ATLAS\textsuperscript{3D} 
& 25.0 & 10.55 & 0.8 & $8.41\pm0.22$ &\citet{Gebhardt_2003_stellar} &2.0
& 0.48 & 2.15 & 10.31
& 0.50 & 2.17 & 10.02 \\
NGC~6861 & MUSE 
& 27.0 & 11.41 & 3.3 & $9.30\pm0.08$ &\citet{Rusli_2013} &3.0
& 0.75 & 1.56 & 10.99
& 0.66 & 1.51 & 10.71 \\
PGC~11179 & PPAK
& 89.0 & 11.07 & 1.8 & $9.28\pm0.08$ &\citet{Cohn_2023_PGC11179} & 3.0
& 0.72 & 1.74 & 10.69
& 0.76 & 1.88 & 10.19 \\
UGC~2698 &  PPAK 
& 91.5 & 11.49 & 3.1 & $9.39\pm0.17$ &\citet{Cohn_Walsh_2021_UGC2698} &3.0
& 0.98 & 1.34 & 11.00
& 1.27 & 0.83 & 10.54 \\
\noalign{\smallskip}
\hline
\end{tabular}
\begin{tablenotes}
\small
\item \textbf{Notes.} Column description: 
\item Column description:
(1) galaxy name;
(2) source of the integral-field unit (IFU) spectroscopic data;
(3) adopted distance $D$ in Mpc;
(4) total dynamical stellar mass $\log(M_{*,\rm dyn}/M_\odot)$
derived from Schwarzschild orbit-superposition modelling in this work;
(5) galaxy effective radius $R_{\rm e}$ in kpc from
\citet{Cappellari_2011_Atlas3D_summarize} for galaxies in the
ATLAS$^{\mathrm{3D}}$ sample, and from \citet{Rusli_2011},
\citet{Emsellem_1999}, and \citet{Rusli_2013} for NGC\,1332,
NGC\,3115, and NGC\,6861, respectively;
(6) black-hole mass $\log(M_{\rm BH}/M_\odot)$ from the literature;
(7) reference for the black-hole mass listed in Column (6);
(8) cut radius $r_{\rm cut}$ in kpc, applied to the time-averaged
radius $r$ of each stellar orbit.

Columns (9)--(11) list the bulge structural parameters for the
$\lambda_z<0.7$ criterion: the bulge effective radius
$R_{\rm e,bulge}$ in kpc, the projected stellar surface-density slope
$\gamma_{\rm bulge}$, and the bulge stellar mass
$\log(M_{\rm bulge}/M_\odot)$. Columns (12)--(14) list the same
quantities for the stricter kinematic criterion $\lambda_z<0.2$.

The black-hole masses follow the same literature sources as
\citet{Graham_2022}, and the original references are listed in
Column (7). Values were rescaled, where necessary, to the distances
in Column (3). The compilation of \citet{Graham_2022} does not include
MRK~1216, NGC~0584, NGC~1271, NGC~1277, NGC~3640, NGC~4281,
NGC~4570, or UGC~2698. For these galaxies, we adopted the additional
measurements cited in Column (7).
\end{tablenotes}
\end{threeparttable}
\end{sidewaystable*}


\section{Results}
\label{sec:analysis}

\subsection{Dynamical bulge vs. photometric bulge}
\label{sec:mass_comparison}

To explore the nature of this dynamical decomposition, we begin by comparing our bulge masses to those obtained from photometric decompositions. Specifically, we use the photometric bulge masses from \citet{Graham_2022}, which provides a large overlap in galaxy samples. Note that our goal is not to reproduce the same bulge mass as the photometric decomposition, since the underlying bulge definitions differ substantially.

As illustrated in Fig.~\ref{fig:bulge_mass_compare2panles}, when the dynamical bulge is defined by $\lambda_z < 0.7$ (panel~a), our bulge masses broadly agree with the photometric bulge masses and follow the one-to-one relation. At the high-mass end, however, our bulge masses become systematically smaller than those of \citet{Graham_2022}, likely because our definition imposes a radial limit of $r_{\rm cut} = 3$ kpc. If we restrict the bulge to only the dynamically hot component with $\lambda_z < 0.2$ (panel~b), the resulting masses are much lower than the photometric bulge masses. This aligns with earlier work indicating that, in massive galaxies, photometrically defined bulges encompass both dynamically hot and warm orbits \citep{Zhu_2018c_model}.

\subsection{Compactness of the bulge}
\label{sec:profiles}

\begin{figure*}
    \centering
    \includegraphics[width=18cm]
        {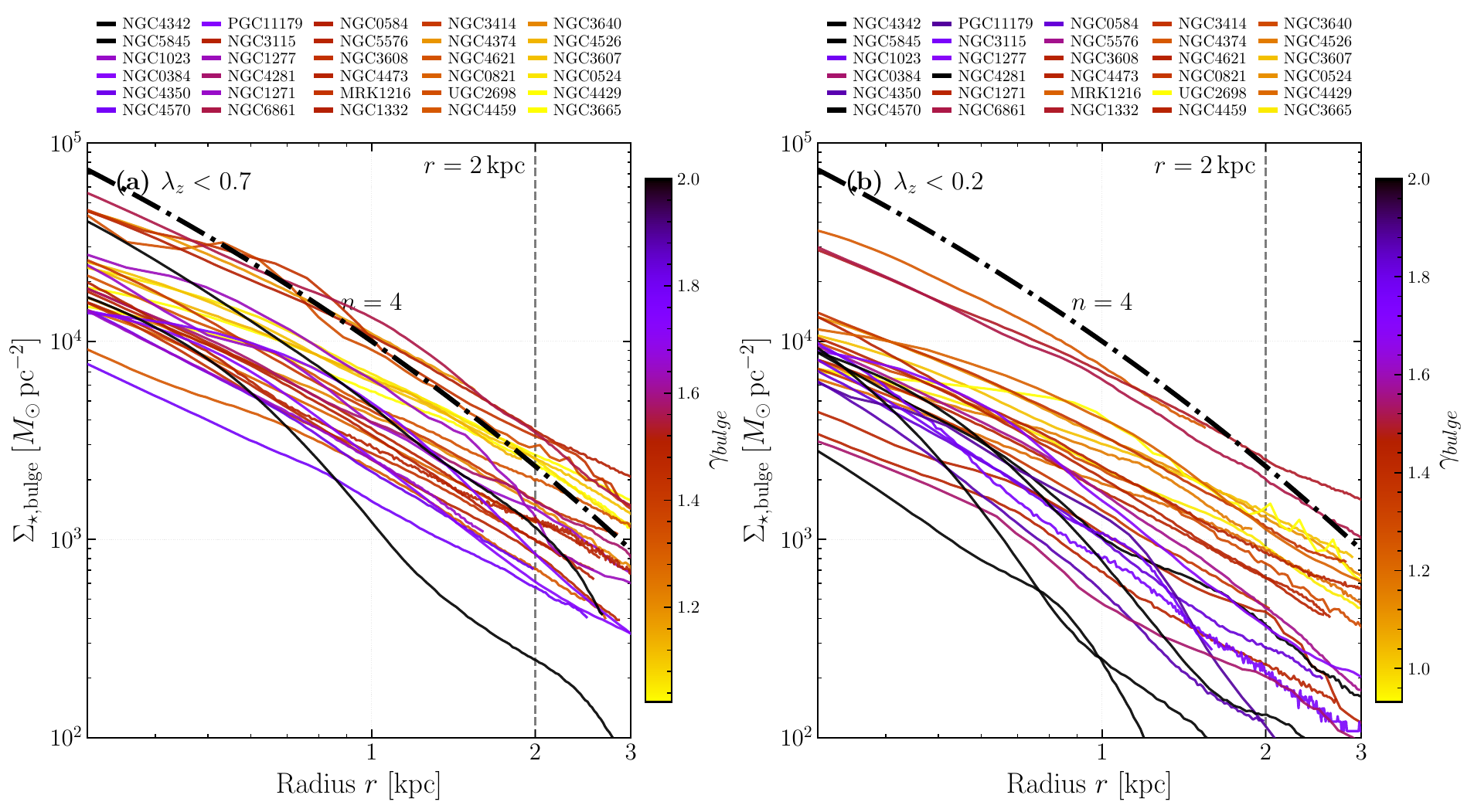}
    \caption{
    Radial profiles of the projected surface mass density of the 
    dynamically defined bulges in the 30 sample galaxies. Panels~(a) and~(b) correspond to bulges defined using $\lambda_z<0.7$ and $\lambda_z<0.2$, respectively.
    Each curve represents one galaxy and is colour-coded by the density slope $\gamma_{\rm bulge}$, as defined in Equation~\ref{func:gamma}. The vertical grey dashed line marks $r=2\,{\rm kpc}$. The dot-dashed curve shows an $n=4$ S\'ersic reference profile with $R_{\rm e}=1\,{\rm kpc}$. 
    }
    \label{fig:surface_density_profiles}
\end{figure*}

We reconstruct the surface mass density on the 2D observational plane of each dynamically defined component from their orbits. We then extract the radial surface mass density
profile, $\Sigma_{\rm bulge}(r)$, along the major axis. The resulting profiles for all 30 galaxies are shown in
\cref{fig:surface_density_profiles}. We show those for bulges with two different
definitions: one with orbits $\lambda_z<0.7$ and one with  $\lambda_z<0.2$.

We introduce two parameters to characterise the compactness of the dynamically defined bulges: the projected bulge effective radius $R_{\rm e,bulge}$ and the surface density slope $\gamma_{\rm bulge}$. The projected half-mass radius $R_{\rm e,bulge}$ is defined as the radius of a circular aperture that contains half of the total projected bulge mass. Recall that we construct the bulge by applying a cut in the time-averaged orbital radius at $r_{\rm cut}$; however, the surface mass density of the bulge assembled from these orbits can still extend beyond $r_{\rm cut}$. While the definition of $R_e$ can therefore still serve as an indicator of bulge compactness, it is nonetheless influenced by this radial cut.

We further define the surface density slope within the data coverage
\begin{equation}
\label{func:gamma}
\gamma_{\rm bulge}=
-\frac{\log \Sigma_{\rm bulge}(2\,{\rm kpc})-
\log \Sigma_{\rm bulge}(0.3\,{\rm kpc})}
{\log(2\,{\rm kpc})-\log(0.3\,{\rm kpc})}.
\end{equation}

Here, $\Sigma_{\rm bulge}(r)$ denotes the mean projected bulge
surface mass density measured within a $0.2$-kpc-wide annulus
centred at radius $r$. Note that the surface mass density used in this analysis has been corrected for the observational point-spread function (PSF) via deconvolution. A larger positive $\gamma_{\rm bulge}$
represents a more rapid decline in surface mass density and
therefore a steeper, more centrally concentrated bulge profile. The values of both parameters are listed in Table~\ref{tab:bulge_properties}.

As shown in Fig.~\ref{fig:surface_density_profiles}, our galaxies cover a wide range in both the absolute surface mass density of the bulge and the shapes of their radial profiles. For bulges—including both dynamically warm and hot orbits, defined by $\lambda_z<0.7$—some galaxies exhibit radial profiles that are broadly consistent with or even steeper than the reference S\'{e}rsic profile with $n=4$, whereas in most systems, the bulge profile is shallower than this $n=4$ reference. In contrast, when the bulge is defined using only dynamically hot orbits ($\lambda_z<0.2$), it shows a lower central density, and its surface density profile is typically shallower than that of the bulge including both warm and hot orbits. A compact warm component is frequently present, particularly in lenticular galaxies. For each galaxy, we therefore compute $R_e$ and $\gamma_{\rm bulge}$ separately for the bulges under these two definitions.The corresponding disc surface mass-density profiles are shown
in \cref{fig:disc_surface_density_profiles_appendix}

\subsection{The $M_{\rm BH}$--$M_{\rm *,bulge}$ scaling relation}
\label{sec:scaling}
With the dynamically defined bulge and bulge compactness characterised. We now investigate where our sample lies on the $M_{\rm BH}$--$M_{\rm *,bulge}$ diagram, and further try to understand the scatter with respect to the bulge compactness, i.e., the bulge effective radius, $R_{\rm e,bulge}$, and the surface-density slope, $\gamma_{\rm bulge}$.

\subsubsection{Colored by $R_{\rm e, bulge}$}
\begin{figure*}
    \centering
    \includegraphics[width=18cm]%
        {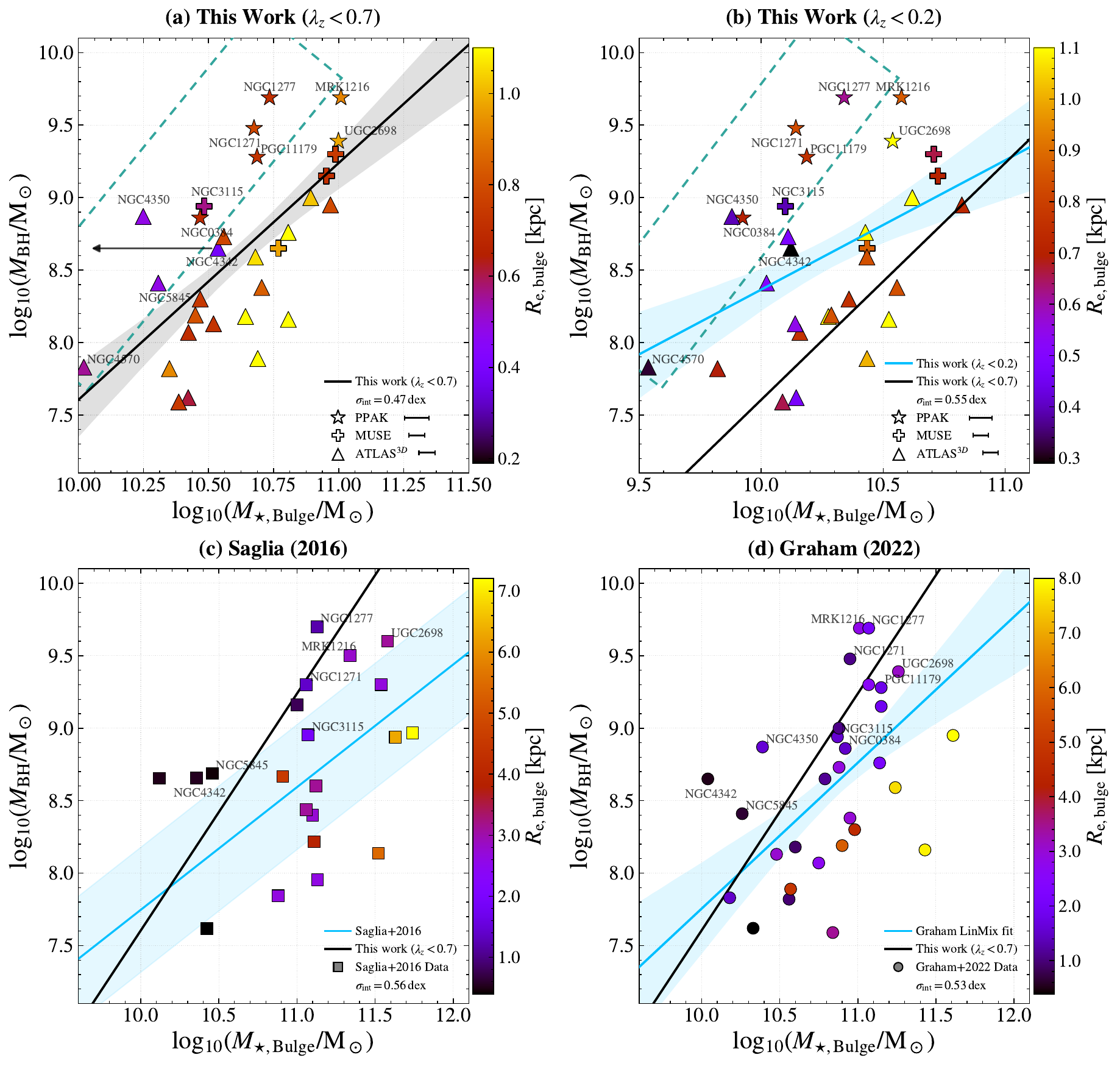}
    \caption{The $M_{\rm BH}$–$M_{\rm *,bulge}$ relations derived using 
dynamical and photometric bulge masses. Individual points are colour-coded 
according to the bulge effective radius $R_{\rm e,bulge}$, 
as indicated by the colour bars. 
\textbf{Panel a:} dynamical bulge masses defined by $\lambda_z<0.7$. Stars, crosses, 
and triangles indicate galaxies with IFU observations from PPAK, 
MUSE, and ATLAS$^{\rm 3D}$, respectively. Typical symmetric $1\sigma$ model uncertainties in $\log_{10}(M_{\rm *,bulge}/M_\odot)$ are shown in the bottom right: 0.047 dex for PPAK, 0.031 dex for MUSE, and 0.032 dex for ATLAS$^{\rm 3D}$. The solid black line 
and shaded regions represent the best-fitting relation and the associated uncertainty. We add an arrow for NGC~4342 pointing to its bulge mass from photometric decomposition \citep{Graham_2022}.
\textbf{Panel b:} same as Panel a but using a dynamical bulge definition of $\lambda_z<0.2$. 
In this panel and also in c, and d, the blue line indicates the corresponding best-fitting relation for the points in each panel, while the black line reproduces the best-fitting relation from Panel a for comparison.
\textbf{Panel c:} photometric bulge masses from \citet{Saglia_2016}, and for MRK\,1216, NGC\,1271, NGC\,1277, and UGC\,2698 from \citet{Walsh_2017_1216h}, \citet{Walsh_2015_1271}, \citet{Walsh_2016_1277} and \citet{Cohn_Walsh_2021_UGC2698}. \textbf{Panel d:} photometric bulge masses from \citet{Graham_2022} and \citet{Savorgnan_2016_bulgemass}, supplemented by measurements from \citet{Cohn_2024_NGC384}, \citet{Cohn_Walsh_2021_UGC2698}, and
\citet{Cohn_2023_PGC11179} for  NGC~0384, UGC~2698, and PGC~11179,
respectively.
The intrinsic scatter around the fitted relation, $\sigma_{\rm int}$, is reported in each panel.
}
\label{fig:MBH_Mbulge_compare_Re}
\end{figure*}

 In \cref{fig:MBH_Mbulge_compare_Re}, we show the $M_{\rm BH}$--$M_{\rm *,bulge}$ relation using our dynamically defined bulges and compare it with relations based on photometrically defined bulges from the literature. Panels~(a) and~(b) show the relations obtained using the 
dynamical bulge masses derived in this work. Panel~(a) uses the bulge defined by 
$\lambda_z<0.7$, whereas panel~(b) uses the bulge defined by $\lambda_z<0.2$. 

We adopt two versions of photometrically defined bulge mass from the literature. 
In Panel~(c), we take the photometrical bulge for most of the galaxies primarily from \citet{Saglia_2016}, while the bulge of MRK\,1216, NGC\,1271, and NGC\,1277 is from \cite{Walsh_2017_1216h}, \cite{Walsh_2015_1271}, \cite{Walsh_2016_1277}, and UGC\,2698 is from \citet{Cohn_Walsh_2021_UGC2698}. In total, 21 of the 30 galaxies are included in this panel. In Panel \textbf{d}, we take the photometric bulge masses from \citet{Graham_2022} and \citet{Savorgnan_2016_bulgemass}, supplemented by measurements from \citet{Cohn_2024_NGC384}, \citet{Cohn_Walsh_2021_UGC2698}, and
\citet{Cohn_2023_PGC11179} for  NGC~0384, UGC~2698, and PGC~11179,
respectively. All 30 galaxies are therefore included.

In each panel, we model the relation using the Bayesian linear-regression code \texttt{LinMix} \citep{Kelly_2007_linmix}. The best-fit parameters for these galaxies, computed for the two alternative dynamical bulge definitions, are reported in Table~\ref{tab:bh_bulge_fit}. When the dynamical bulge is defined by $\lambda_z<0.7$ and $r< r_{\rm cut}$, we obtain a steep slope of 1.63 in the $M_{\rm BH}-M_{\star, \rm bulge}$ relation (see Table~\ref{tab:bh_bulge_fit}), which is slightly steeper than the relation derived for ES/S0 galaxies in \citet{Graham_2022}. With the bulge defined by $\lambda_z<0.2$ and $r<r_{\rm cut}$, the scatter becomes larger, and the fitting results yield a slope of 0.91.

In our regression analysis, $\sigma_{\rm int}$ denotes the intrinsic scatter in
$\log_{10}(M_{\rm BH}/M_\odot)$ after accounting for the adopted
measurement uncertainties. It is measured at fixed
$M_{\star,\rm bulge}$ for the univariate relations and at fixed $M_{\star,\rm bulge}$ and either $\log_{10}(R_{e,\rm bulge}/{\rm kpc})$ or
$\gamma_{\rm bulge}$ for the bivariate relations.
We report the posterior median, with the lower and upper
uncertainties determined from the 16th and 84th percentiles,
respectively.
The intrinsic scatter $\sigma_{\rm int}$ around the best-fitting model line is 0.47 and 0.55 dex in panels a and b for the dynamically defined bulge, whereas it increases to 0.56 and 0.53 dex in panels c and d when using the two photometrically defined bulge definitions. 
With the dynamically defined bulge—particularly when the bulge is defined by $\lambda_z<0.7$—the correlation is noticeably tighter.

\begin{figure*}
    \centering
    \includegraphics[width=18cm]{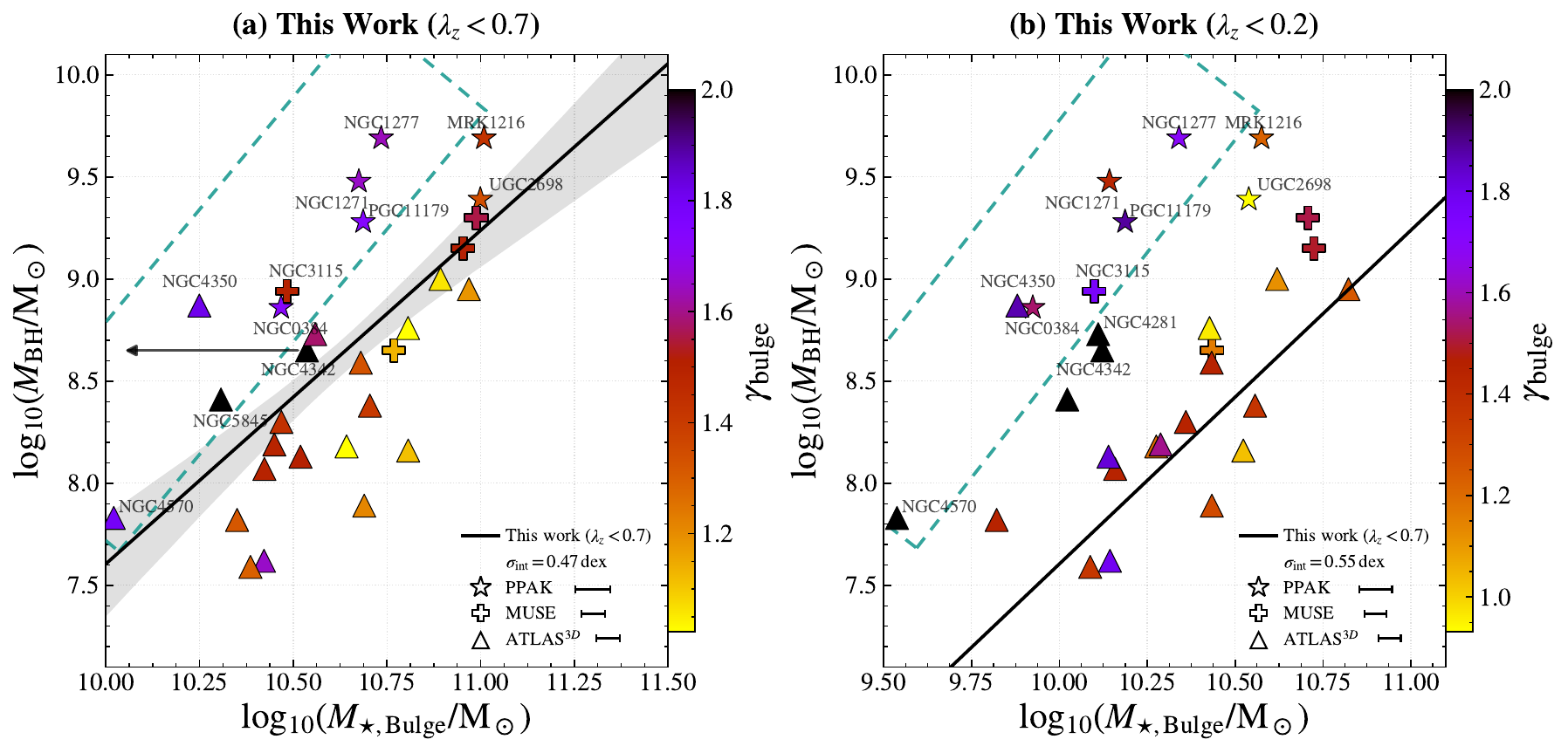}
    \caption{\textbf{Same as Fig.~\ref{fig:MBH_Mbulge_compare_Re}, but with the data points colour-coded by the bulge surface density slope $\gamma_{\rm bulge}$ defined in Equation~\ref{func:gamma}.} At fixed bulge mass, galaxies hosting overmassive BHs, selected by a dashed boxlet, have the steepest bulge density slope (large $\gamma_{\rm bulge}$).
    }
    \label{fig:MBH_Mbulge_compare_gamma}
\end{figure*}

The four panels of \cref{fig:MBH_Mbulge_compare_Re} are colour-coded according to the bulge effective radius. Panels~(a) and~(b) adopt the $R_{\rm e,bulge}$ defined for the dynamical bulge, as introduced in Section \ref{sec:profiles}, while panels~(c) and~(d) use the bulge $R_e$ from the photometric decomposition.

In Panel~(a), we identify a subset of galaxies—NGC~4570, NGC~5845, NGC~4350, NGC~0384, NGC~3115, PGC~11179, NGC~1277, and NGC~1271, included in the dashed boxlet—that host overly massive BHs. They appear to form a roughly parallel sequence relative to the rest of the sample. These systems have the smallest $R_{\rm e,bulge}$ at fixed bulge mass, with a mean $R_{\rm e,bulge}=0.62\,{\rm kpc}$ compared to $0.88\,{\rm kpc}$ for the other galaxies. Likewise, in Panel~(b), the circled galaxies with overmassive BHs have an average $R_{\rm e,bulge}=0.56\,{\rm kpc}$ versus $0.81\,{\rm kpc}$ for the remaining objects. At a given bulge mass, galaxies with more compact bulges tend to host more massive BHs.

When the bulge is instead defined photometrically, as in panels~(c) and~(d), a similar pattern is suggested: points lying above the best-fitting relation generally correspond to smaller bulge sizes. However, the scatter is larger.

\subsubsection{Colored by $\gamma_{\rm bulge}$} 
We further test this dependence by colour-coding the points according to the surface density slope $\gamma_{\rm bulge}$ of the dynamically defined bulge. The $M_{\rm BH}$--$M_{\rm *,bulge}$ relations for the two versions of dynamical bulges are presented in \cref{fig:MBH_Mbulge_compare_gamma}. We find a similar pattern indicating that galaxies hosting overmassive BHs tend to live in bulges with the steepest surface density profiles. 

When the bulge is defined to include both dynamically warm and hot orbits ($\lambda_z<0.7$), the set of galaxies with overmassive BHs, as selected by the boxlet, has an average $\overline{\gamma_{\rm bulge}}=1.75$, compared to $1.38$ for the rest of the sample. If we restrict the bulge to orbits with $\lambda_z<0.2$, the galaxies with overmassive BHs have $\overline{\gamma_{\rm bulge}}=1.85$, while the others have $1.45$. At fixed bulge mass, galaxies with overmassive BHs, therefore, exhibit the steepest bulge surface density profiles.

One exceptional galaxy may be NGC~4342. It hosts a very compact bulge; however, it does not appear to contain a strongly overmassive BH in our relation when defining bulge with $\lambda_z<0.7$. As already shown in Fig.~\ref{fig:bulge_mass_compare2panles}, NGC~4342 is the only galaxy for which our dynamical bulge mass is substantially higher than the photometric bulge mass. From the image, NGC~4342 resembles a disc galaxy with only a small bulge, but it contains a significant fraction of warm orbits ($\lambda_z \sim 0.5$) that were assigned to the bulge under our definition of $\lambda_z<0.7$. These warm orbits were likely heated by the external environment (e.g., tidal forces, \citealp{Blom_2014_NGC4342}), given the galaxy's position in the Virgo cluster. It may be in a phase where the disc has already been significantly heated into warm components while the central bulge remains compact and largely unaffected. We, therefore, added an arrow indicating its bulge mass from the photometric decomposition. It moves to the region with overmassive BHs when the bulge is defined by $\lambda_z<0.2$ as shown in the right panel.

\begin{figure*}
    \centering
    \includegraphics[width=16cm]{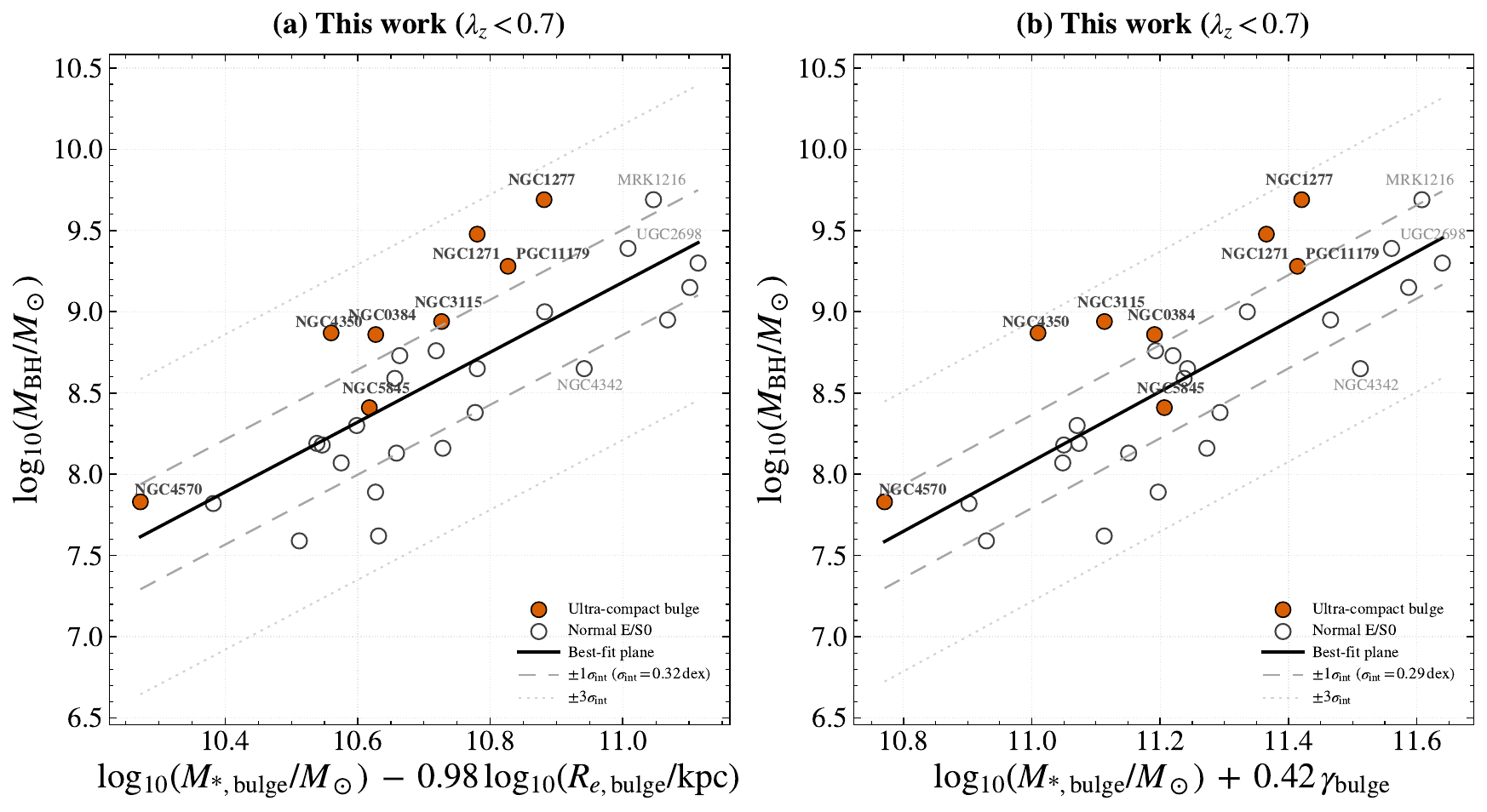}
    \caption{\textbf{$M_{\rm BH}-M_{\rm bulge}-R_{\rm e, bulge}$ (left) and $M_{\rm BH}-M_{\rm bulge}-\gamma_{\rm bulge}$ (right) relation.} By introducing a secondary parameter describing the bulge compactness ($R_{\rm e, bulge}$ or $\gamma_{\rm bulge}$), the intrinsic scatter around the best-fitting relation is reduced to $\sim 0.3$.  }
    \label{fig:Mbh-Mbulge-gamma}
\end{figure*}

\subsection{Secondary dependence on bulge compactness}
\label{sec:structure_dependence}
We take the dynamical bulge mass defined by $\lambda_z<0.7$ in the following analysis. We consider the compactness of the bulge, either $R_{\rm e, bulge}$ or $\gamma_{\rm bulge}$ as the secondary parameter, and refit the BH-bulge relation using the following equation:
\begin{equation}
\log_{10}\left(\frac{M_{\rm BH}}{M_\odot}\right)
=
\alpha
 [\log_{10}\left(\frac{M_{\ast,\rm bulge}}{M_\odot}\right)
+\eta Y]+\beta,
\qquad \lambda_z<0.7,
\label{eq:mbh-bulge-plane}
\end{equation}
Here, $Y=\log_{10}(R_{e,\rm bulge}/{\rm kpc})$ for the
bulge-mass--size plane and $Y=\gamma_{\rm bulge}$ for the
bulge-mass--slope plane. The coefficient $\alpha$ describes
the dependence on bulge stellar mass, $\eta$ describes the
dependence on the secondary structural parameter, and $\beta$
is the intercept of the uncentered relation.

We take the dynamical bulge defined by $\lambda_z<0.7$, as shown in Fig.~\ref{fig:Mbh-Mbulge-gamma}; the relation becomes tighter by including the secondary parameter. The scatter around the best-fitting line is $\sigma_{\rm int} = 0.32$ for the $M_{\rm BH}-M_{\rm bulge}-R_{\rm e, bulge}$ relation and $\sigma_{\rm int} = 0.29$ for $M_{\rm BH}-M_{\rm bulge}-\gamma_{\rm bulge}$, which is significantly smaller than the original $\sigma_{\rm int} = 0.47$ without the secondary parameter on bulge compactness. The best-fitting values of $\alpha$, $\eta$, and $\beta$,
together with the intrinsic scatter $\sigma_{\rm int}$, are
listed in Table~\ref{tab:bh_bulge_fit}.

Note that once we include $R_{\rm e, bulge}$, we find $M_{\rm BH} \sim (M_{*,\rm bulge}/R_{\rm e, bulge}^{0.98})^{2.15}$, which is broadly consistent with $M_{\rm BH} \sim (M_{*,\rm bulge}/R_{\rm e, bulge})^{2.15} \sim \sigma_{\rm bulge}^{4.30}$. Considering bulges approximately satisfy the virial equilibrium $\sigma_{\rm bulge}^2 \sim GM_{\rm bulge}/R_{\rm e, bulge}$, the scaling relation we derived is essentially an alternative expression of the $M_{\rm BH}-\sigma$ relation. This agrees with the observation that the outliers we highlight in the $M_{\rm BH}- M_{\rm bulge}$ relation still lie on the $M_{\rm BH}-\sigma$ relation; NGC 1277, NGC 1271, NGC 0384, PGC 11179, and MRK~1216 are shown in \citep{Cohn_2024_NGC384}; NGC 4570 is shown in \cite{Thater_2019}; NGC 4350 and NGC 3115 are included in \citep{Savorgnan2015MNRAS.446.2330S}; and NGC 4342 and NGC 5845 appear in \citep{Saglia_2016}. 

The $M_{\rm BH}-M_{\rm bulge}-R_{\rm e, bulge}$ relation obtained here resembles that reported by \citep{van_den_Bosch_2016} for a general galaxy sample, where BH mass is linked to the total galaxy mass and size. In both cases, the physical explanation is tied to the virial theory of the stellar system, i.e., the fundamental plane. However, we have been able to focus on the bulge, which results in significant smaller scatter than the relation reported by \citep{van_den_Bosch_2016} , within this work, we specifically unify the distribution of ETGs.

\begin{table*}[htbp!]
\caption{
Linear and bivariate regression results for the black hole
mass scaling relations.
}
\label{tab:bh_bulge_fit}
\centering
\small
\renewcommand{\arraystretch}{1.15}
\begin{tabular}{l c c c c c}
\hline\hline

Sample &
$N$ &
Mass slope &
Secondary slope &
Intercept &
$\sigma_{\rm int}$ \\
&
&
$\alpha$ &
$\eta$ &
$\beta$ &
(dex) \\

\hline

\multicolumn{6}{c}{
$\displaystyle
\log_{10}\left(\frac{M_{\rm BH}}{M_\odot}\right)
=
\alpha
\log_{10}\left(
\frac{M_{\ast,\rm bulge}}
     {5\times10^{10}\,M_\odot}
\right)
+\beta$
}
\\

\hline

This work ($\lambda_z<0.7$)
&
30
&
$1.63\pm0.36$
&
\ldots
&
$8.73\pm0.09$
&
$0.47\pm0.07$
\\
\thinhline

This work ($\lambda_z<0.2$)
&
30
&
$0.91\pm0.37$
&
\ldots
&
$8.99\pm0.19$
&
$0.55\pm0.08$
\\
\thinhline

\begin{tabular}[c]{@{}l@{}}
This work ($\lambda_z<0.7$)\\[-2pt]
E/S0 with ultra-compact bulges
\end{tabular}
&
8
&
$2.36\pm0.47$
&
\ldots
&
$9.45\pm0.17$
&
$0.19\pm0.13$
\\
\thinhline

\begin{tabular}[c]{@{}l@{}}
This work ($\lambda_z<0.7$)\\[-2pt]
normal E/S0 galaxies
\end{tabular}
&
22
&
$2.23\pm0.31$
&
\ldots
&
$8.55\pm0.06$
&
$0.27\pm0.06$
\\
\thinhline

Graham et al.\ (2022; ES/S0)\\[-2pt]
\texttt{LinMix} fit
&
30
&
$1.01\pm0.35$
&
\ldots
&
$8.46\pm0.12$
&
$0.53\pm0.09$
\\
\thinhline

Graham et al.\ (2022; ES/S0)
&
40
&
$1.53\pm0.15$
&
\ldots
&
$8.57\pm0.15$
&
\ldots
\\

\hline\hline

\multicolumn{6}{c}{
$\displaystyle
\log_{10}\left(\frac{M_{\rm BH}}{M_\odot}\right)
=
\alpha
[\log_{10}\left(\frac{M_{\ast,\rm bulge}}{M_\odot}\right)
+\eta Y]+\beta$
}
\\

\hline

\begin{tabular}[c]{@{}l@{}}
This work ($\lambda_z<0.7$)\\[-2pt]
$Y=\log_{10}(R_{e,\rm bulge}/{\rm kpc})$
\end{tabular}
&
30
&
$2.15\pm0.27$
&
$-0.98\pm0.32$
&
$-14.48\pm2.92$
&
$0.32\pm0.05$
\\
\thinhline

\begin{tabular}[c]{@{}l@{}}
This work ($\lambda_z<0.7$)\\[-2pt]
$Y=\gamma_{\rm bulge}$
\end{tabular}
&
30
&
$2.15\pm0.25$
&
$0.42\pm0.12$
&
$-15.57\pm2.74$
&
$0.29\pm0.06$
\\

\hline
\end{tabular}

\tablefoot{
The first section presents the single-variable
$M_{\rm BH}$--$M_{\ast,\rm bulge}$ relations, written as
$\log_{10}(M_{\rm BH}/M_\odot)
=
\alpha
\log_{10}[M_{\ast,\rm bulge}/(5\times10^{10}\,M_\odot)]
+\beta$.
The second section presents the bivariate relations, written as
$\log_{10}(M_{\rm BH}/M_\odot)
=
\alpha\left[
\log_{10}(M_{\ast,\rm bulge}/M_\odot)
+\eta Y
\right]+\beta$,
where $Y$ is either
$\log_{10}(R_{e,\rm bulge}/{\rm kpc})$
or $\gamma_{\rm bulge}$. Here, $\eta$ is the secondary
coefficient relative to the bulge-mass slope, such that
$\alpha\eta$ gives the direct coefficient of $Y$.
Here, $N$ is the number of galaxies, and
$\sigma_{\rm int}$ is the intrinsic scatter in the
$\log M_{\rm BH}$ direction.
The parameters for this work were derived using the
\citet{Kelly_2007_linmix} Bayesian regression method.
Ellipses indicate parameters that are not applicable or
are not reported for the corresponding fit.
}

\end{table*}

\begin{figure*}
    \centering
    \includegraphics[width=11cm]{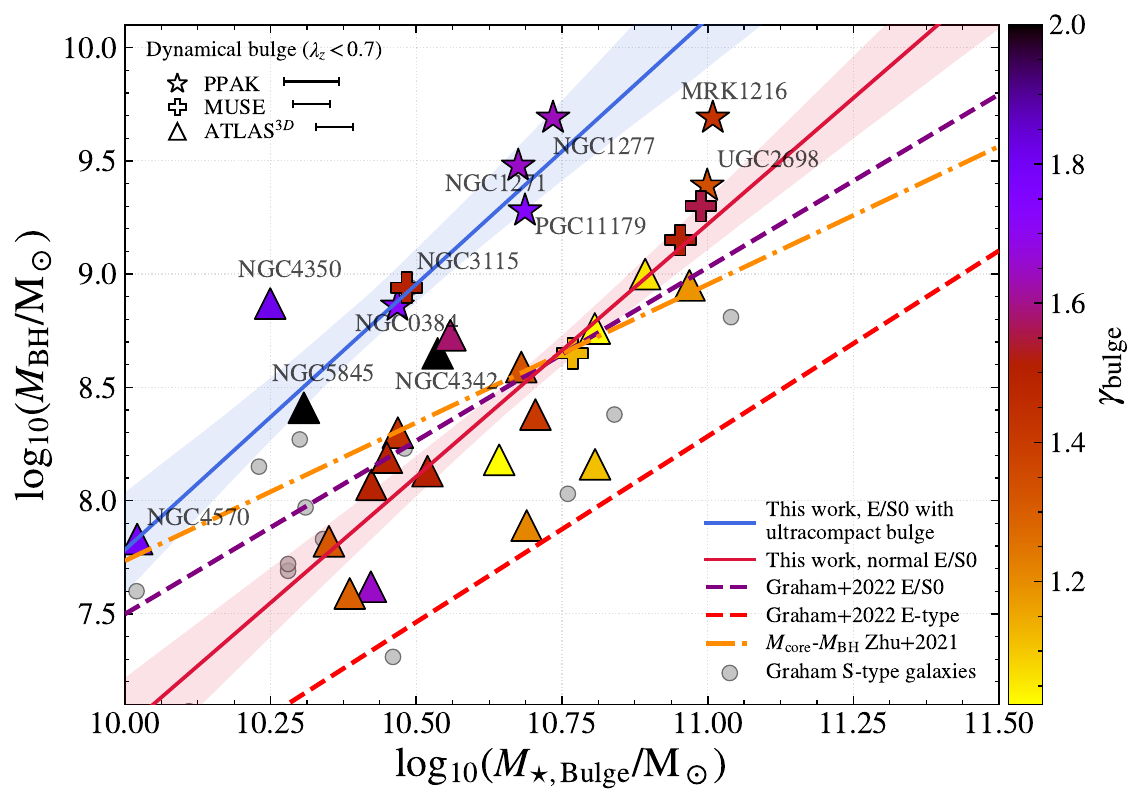}
    \caption{\textbf{Evolution of the $M_{\rm BH}-M_{*,\rm bulge}$ relation}. With bulge dynamically defined by orbits with $\lambda_z<0.7$, the majority of our galaxies define a relation (red solid line) that is broadly consistent with the E/S0 relation reported by \citet{Graham_2022} (purple dashed line) and with the $M_{\rm BH}-M_{*,\rm core}$ relation from \citet{Zhu&Ho2021ApJ...907....6Z} (orange dot-dashed line). By comparison, the ultracompact-bulge systems follow an approximately parallel trend (blue solid line), but with BH masses elevated by about one dex at a given bulge mass. Conversely, the giant ellipticals in \citet{Graham_2022} (red dashed) show BH masses that are roughly one dex lower. Our major sample yields a steeper relation than previous studies, the spiral galaxies from \citet{Graham_2022} (gray dots) are consistent with our relation at the low mass end.}
    \label{fig:compare}
\end{figure*}

\section{Discussion}
\label{sec:discussion}

\subsection{Evolution of $M_{\rm BH}$--$M_{\rm bulge}$ relation}
\label{sec:discussion_compactness}
The dependence of the $M_{\rm BH}$--$M_{\rm bulge}$ on the bulge compactness may indicate an evolution of the relation with galaxy assembly history. We discuss the implications combined with previous results from photometric decompositions.

In Fig.~\ref{fig:compare}, we fitted two nearly parallel sequences in the $M_{\rm BH}$–$M_{\rm bulge}$ plane, one traced by the majority of our sample and the other by the eight outliers: NGC 4570, NGC 5845, NGC 4350, NGC 0384, NGC 3115, PGC 11179, NGC 1271, and NGC 1277. 
The best-fit relation for the main population broadly agrees with that reported by \citet{Graham_2022} for E/S0 systems, as well as with the relation in \citet{Zhu&Ho2021ApJ...907....6Z}, where the bulge mass is defined photometrically but using the inner core regions. This agreement is expected because our galaxies are predominantly E/S0, as classified by \citet{Graham_2022};  we excluded giant ellipticals since the IFU coverage is limited relative to their large bulge sizes. Nevertheless, our fitted relation is steeper (i.e., has a higher slope), likely because our adopted radial cut for defining the bulge preferentially reduces the inferred bulge mass for the largest bulges in the most massive galaxies. With a steeper relation, the spiral galaxies from \citet{Graham_2022} become consistent with our relation at the low-mass end. Although we still need a dynamically defined bulge for spiral galaxies for a fair comparison. 

The eight outliers, which host extremely compact bulges, follow a separate sequence with BH masses roughly one dex higher than those of the main population. 
All eight of our outliers are classified as (ultra)compact galaxies with small overall sizes ($R_{\rm e,bulge}=0.48$--$0.78\,\mathrm{kpc}$ for the $\lambda_z<0.7$ definition; see Table~\ref{tab:bulge_properties}). Yet, they retain discs that are relatively extended compared to their very compact bulges. If mergers destroy these discs, the resulting increase in bulge mass and size could shift the remnants onto the normal S0+ES sequence.

In contrast, \citet{Graham_2022} derived a relation for giant ellipticals with the most massive bulges, whose BH masses are about one dex lower.
\citet{Graham_2022} argued that the evolution of the $M_{\rm BH}$–$M_{\rm bulge}$ relation is linked to a galaxy’s merger history. In their picture, dry mergers drive the transition from E/S0 galaxies to giant ellipticals by increasing spheroid mass more efficiently than BH mass, both by roughly doubling the bulge mass and by disrupting the progenitor discs. 

Taken together, galaxies with extremely compact bulges, normal S0+ES galaxies, and giant ellipticals may represent a hierarchical, merger history-driven evolution of the $M_{\rm BH}$–$M_{\rm bulge}$ relation.
Such evolution does not seem to strongly depend on the current environments of the galaxies: the eight outliers we identified include NGC 4570 and NGC 4350 in the Virgo cluster, an isolated galaxy, NGC 3115, and the remaining five in groups.

High-redshift observations indicate that black holes can be over-massive by $\sim 10-100 \times$
\citep{Pacucci2023ApJ...957L...3P,Maiolino2024Natur.627...59M}, or even more in the so-called “Little Red Dots” \citep{Durodola2025ApJ...985..169D}, relative to low-$z$ galaxies with hosts of the same stellar mass. 
 The Green Pea galaxies, serving as local relics of the little red dots, also show consistently overmassive BH \citep{Lin2025ApJ...980L..34L}. This observational evidence establishes a scenario involving the evolution of the $M_{\rm BH}-M_{\rm bulge}$ relation with redshift. All these studies are based on AGN-based BH mass measurements, and most BHs in little red dots and Green Pea Galaxies have $M_{\rm BH} \lesssim 10^{8}$\Msun, a regime in which local and high-redshift galaxies differ markedly. While AGN-based measurements suggest that the local–high-$z$ difference is no longer clear \citep{Lin2025ApJ...980L..34L}, this may be due to the large uncertainty in the BH mass. The BH mass measured directly from kinematics at high-z is only available for a few lensed systems\citep{Newman2026Sci...392.1065N,Juodvzbalis2026Natur.653.1017J}.

With BH mass measured by stellar kinematics at a similar accuracy as that of other local galaxies, systems with exceptionally quiescent merger histories, such as the eight outliers in this work, may therefore provide a key intermediate population, bridging local galaxies and their high-redshift counterparts.
The outliers in our study lie in the high-mass range with $M_{\rm BH} \gtrsim 10^{8}$\Msun, thus providing the unique chance to imply that local and high-$z$ galaxies still differ at the massive end.

\subsection{Not all relic-galaxy candidates are truly merger-free}
\label{sec:relic-candidates}
Our sample comprises six ultra-compact massive ETGs that have been identified as relic-galaxy candidates \citep{Y_ld_r_m_2017}: MRK0~1216, NGC~0384, NGC~1271, NGC~1277, PGC~11179, and UGC2698. Some of these galaxies have previously been argued to host overmassive BHs \citep{Walsh_2015_1271,Walsh_2017_1216h}, though with considerable scatter \citep{Cohn_2024_NGC384}; other studies have instead reached the more debated conclusion that their BH masses are broadly typical \citep{Savorgnan_2016_bulgemass}. Our findings indicate that NGC~1277, NGC~1271, PGC~11179, and NGC~0384 indeed contain strongly overmassive BHs. MRK~1216 also appears mildly overmassive. By contrast, UGC~2698 lies on our $M_{\rm BH}$–$M_{\rm bulge}$ relation, consistent overall with the photometric decomposition analysis \citep{Cohn_2024_NGC384}.

We emphasise that these relic candidates span a broad diversity of internal dynamical configurations and that not all of them can be viewed as truly merger-free systems. Although the six galaxies considered here have comparable global sizes, NGC~1277, NGC~1271, PGC~11179, and NGC~0384 show relatively extended, dynamically cold discs together with extremely compact bulges. In contrast, MRK~1216 and UGC~2698 show less cold disc and have more extended bulges; their internal structure was likely reshaped by additional dynamical mechanisms or modest mergers, which would have increased both bulge mass and bulge size. The compactness of the dynamically hot bulge is a powerful indicator of merger history: galaxies with extremely compact bulges tend to have exceptionally quiescent merger histories \citep{Zhu_2025_merger_free}. In \citet{Zhu_2025_merger_free}, the merger histories inferred from bulge compactness for these six systems were ordered as NGC~1277, PGC~11179, NGC~0384, NGC~1271, MRK~1216, and UGC~2698, ranging from extremely quiet merger histories to possibly moderate merger activity.

From this perspective, we interpret the spread of relic-galaxy candidates on the $M_{\rm BH}$–$M_{\rm bulge}$ relation as a consequence of their differing merger histories. The most merger-free relics—flagged by their ultra-compact bulges—host overly massive BHs; the clearest examples are NGC~1277 and PGC~11179, with NGC~1271 and NGC~0384 being slightly less compact. Conversely, UGC~2698 has a comparatively large bulge, suggesting that moderate mergers or other dynamical processes have altered its structure, and it accordingly falls close to our standard $M_{\rm BH}$–$M_{\rm bulge}$ relation. MRK~1216 appears intermediate, both in bulge compactness and in the degree to which its BH is overmassive.

\subsection{The dynamically defined bulge and the Effect of $r_{cut}$}
\label{sec:rcut_bulge_mass}

We apply a fixed radial threshold, \(r_{\rm cut}\), to define the bulge as the predominantly bound, dynamically hot/warm orbital component. We emphasise that this cut is based on each orbit’s time-averaged radius—which scales with binding energy—rather than the particles’ instantaneous radii. Our default choice is \(r_{\rm cut}=3\,\mathrm{kpc}\); however, for a few galaxies, the IFU data do not extend out to this distance, so we instead adopt \(r_{\rm cut}=2\,\mathrm{kpc}\).

This limitation is unlikely to substantially affect the bulge mass of NGC~4342, NGC~4350, and NGC~5845. These galaxies are extremely compact, with $R_e< 1.5$ kpc. From their stellar orbit distribution, there is clear evidence that their extended stellar light is a dynamically cold disc. The bulge has already been fully covered within 2 kpc; extending to 3 kpc will not add anything to their bulge mass.

Taking $r_{\rm{cut}}=2\,\mathrm{kpc}$ may cause a systematic underestimation of the bulge mass for NGC~0584, NGC~1023, NGC~3608, NGC~4281,NGC~4459, NGC~4473, NGC~4526, NGC~4570, NGC~4621, NGC~5576. 
Note that all these galaxies lie on the main sequence of the $M_{\star, \rm bulge}-M_{\rm BH}$ relation with the current bulge mass; none of them are identified as outliers with overly massive BHs. If we consider their bulge mass increase $30\%$ by allowing $r_{\rm{cut}}=3\,\mathrm{kpc}$, they will move rightward by about 0.1 dex without affecting our major conclusion.

It is also important to note that there are certainties in the stellar orbit distribution we obtained, depending on the data quality. As we learnt from the model test against mock data \citep{Zhu_2020}, if a galaxy has a compact bulge and an extended disc, the model might introduce a small fraction of dynamically hot orbits in the outer regions, which will cause the bulge to be less compact and over-estimate the bulge mass if we include those orbits. For instance, the small bump of hot orbits at $r\sim 4$ kpc for NGC 1277 is likely model noise (See \Cref{fig:circularity25}). Taking a small $r_{\rm{cut}}$ for galaxies with ultra-compact bulges could help to suppress such contamination. 

\section{Conclusion}
\label{sec:conclusion}
We compile a sample of 30 early-type galaxies 
with black hole masses measured dynamically from spatially resolved stellar or gas kinematics and wide-field IFU data from the literature. Giant ellipticals with a central core are excluded, the sample is dominated by S0 and elliptical galaxies still with steep surface density in the inner regions. For each galaxy, we construct an orbit-superposition Schwarzschild model to infer the intrinsic stellar orbital configuration and perform a dynamical separation into bulge and disc components.

We consider two different bulge definitions. In the first, the bulge comprises dynamically warm and hot orbits with circularity $\lambda_z < 0.7$ within $r < 3$ kpc. In the second, the bulge is limited to dynamically hot orbits only, requiring $\lambda_z < 0.2$ and $r < 3$ kpc. In every case, the bulge is restricted to the compact central region ($r < 3$ kpc), where $r$ is the time-averaged orbital radius scaling with the orbital energy. For each definition, we rebuild the bulge surface-density distribution from the corresponding orbits and characterise its compactness using two quantities: the effective radius $R_\mathrm{e,bulge}$ and the inner surface-density slope $\gamma_{\rm bulge}$ measured within 2 kpc.
    
Our main conclusions are as follows:
\begin{enumerate}

  \item Using the bulge definition with $\lambda_z < 0.7$ and $r<3$ kpc, which encompasses both dynamically hot and warm orbits, the inferred dynamical bulge masses are, in general, consistent with the photometric bulge masses reported in the literature. In contrast, the bulges restricted to only dynamically hot orbits ($\lambda_z < 0.2$) are systematically lower in mass.

\item Dynamical bulge masses computed with $\lambda_z < 0.7$ produce a tighter $M_\mathrm{BH}$--$M_{\rm bulge}$ relation, with a scatter of $\mathcal{\sigma} = 0.47$ dex around the best-fitting line for the 30 galaxies compared with the relation based on photometric bulges from literature decompositions ($\sigma = 0.55$ dex). By comparison, defining the bulge using only dynamically hot orbits ($\lambda_z < 0.2$) results in a larger scatter in the $M_\mathrm{BH}$--$M_{\rm bulge}$ relation, $\sigma = 0.55$.

\item With this dynamically defined bulge, we find a clear pattern: at fixed dynamical bulge mass, galaxies with more compact bulges—i.e., smaller effective radii $R_{\rm e, bulge}$ and/or steeper surface-density slopes $\gamma_{\rm bulge}$—preferentially host more massive BHs. The eight outliers, NGC~4570, NGC~5845, NGC~4350, NGC~0384, NGC~3115, PGC~11179, NGC~1271, and NGC~1277, lie on a distinct sequence, with BH masses roughly one dex higher than those of the main galaxy population in our sample. The pattern indicates the evolution of $M_\mathrm{BH}$--$M_{\rm bulge}$ with galaxy merger history; the eight outliers are likely with extremely quiescent merger history encoded in their exceptionally compact bulges.

\item Taking the bulge defined by $\lambda_z<0.7$, incorporating the bulge compactness via an $M_{\rm BH}-M_{\rm *, Bulge}-R_{\rm e, bulge}$ (or $\gamma_{\rm bulge}$) relation brings the outliers into agreement and reduces the overall scatter of the sample to $\sim 0.3$ dex. The $M_{\rm BH}-M_{\rm *, Bulge}-R_{\rm e, bulge}$ relation presents as an alternative of the $M_{\rm BH}-\sigma_{\rm bulge}$ relation considering the bulge satisfy the virial equilibrium. 
\end{enumerate}

The combination of IFU observations and the orbit-superposition model enables us to dynamically define a bulge for real galaxies.  
Our results indicate that the $M_\mathrm{BH}$–$M_{\rm bulge}$ relation evolves with a galaxy’s merger history. Future combination of stellar populations \citep{Zhu_2020} will help us to define more decisive in-situ bulges. Applying this dynamical decomposition to barred galaxies \citep{Tahmase2022ApJ...941..109T, Jin2026arXiv260700497J} will be our next step toward clarifying black hole–galaxy coevolution in barred systems.
    
\begin{acknowledgement}
The research presented here is supported by the NSFC key project 12633002 (LZ), National Key R\&D Programme of China No. 2025YFF0511002 (LZ), and CAS Project for Young Scientists in Basic Research, Grant No. YSBR-062 (LZ). ACS acknowledges support from FAPERGS (grants 23/2551-0001832-2 and 24/2551-0001548-5), CNPq (grants 314301/2021-6, 445231/2024-6, 404233/2024-4 and 312940/2025-4), and CAPES (grant 88887.004427/2024-00).
\end{acknowledgement}

\bibliographystyle{aa}  
\bibliography{bulge} 

\begin{appendix}

\section{Model related figures}
\label{app:mge}

The surface brightness of NGC\,1332 was modelled using the 
Multi-Gaussian Expansion \citep[MGE;][]{Emsellem_1994_MGE} method, 
as implemented in the \textsc{MgeFit} package of 
\citet{Cappellari_2002_MGE}. 
We use an archival 
image from the DESI Legacy Survey \citep{Dey_2019_DESI} in the 
$r$ band, with a pixel scale of $0\farcs262$\,pixel$^{-1}$.
The point spread function (PSF) is modelled as a single circular 
Gaussian with $\mathrm{FWHM} = 1\farcs0$, consistent with the 
typical seeing conditions of the DESI Legacy Survey.

Prior to fitting, the sky background is estimated from the image 
corners using iterative $2.5\sigma$-clipping and subtracted.
Foreground stars and background sources are masked and replaced 
by the local median of a surrounding annulus.
We then fit the sky-subtracted image using the \textsc{MgeFit} 
package \citep{Cappellari_2002_MGE}, following the standard 
procedure of \citet{Scott_2013_ATLAS_MGE}: the galaxy geometry 
(centre, position angle $\mathrm{PA} = 114\fdg6$, and ellipticity 
$\varepsilon = 0.66$) is first determined with the 
\textsc{find\_galaxy} routine, after which 
\textsc{sectors\_photometry} extract the azimuthally averaged 
surface brightness profile along 19 equally spaced sectors.
The MGE fit is performed with \textsc{mge\_fit\_sectors}, using 
a minimum surface brightness threshold of 
$10\,L_\odot\,\mathrm{pc}^{-2}$ (approximately $3\times$ the 
sky background level) to exclude sky-noise-dominated pixels at 
large radii.

The MGE parameters are listed in Table~\ref{tab:mge_ngc1332}.

\begin{table}
\caption{MGE parameters for NGC\,1332.}
\label{tab:mge_ngc1332}
\centering
\begin{tabular}{r r r}
\hline\hline
\noalign{\smallskip}
$I_j$ & $\sigma_j$ & $q_j$ \\
(L$_\odot$/pc$^2$) & (arcsec) & \\
(1) & (2) & (3) \\
\noalign{\smallskip}
\hline
\noalign{\smallskip}
25360.52 & 0.38  & 0.916 \\
11133.92 & 1.11  & 1.000 \\
14318.22 & 2.12  & 0.831 \\
 6218.96 & 4.13  & 0.742 \\
 3500.31 & 7.56  & 0.814 \\
 1017.66 & 15.91 & 0.348 \\
  791.33 & 31.29 & 0.284 \\
  347.21 & 48.80 & 0.335 \\
    8.02 & 87.50 & 1.000 \\
   59.04 & 87.50 & 0.437 \\
\noalign{\smallskip}
\hline
\end{tabular}
\tablefoot{
Columns: (1) central surface brightness of each Gaussian component; 
(2) Gaussian standard deviation along the major axis; 
(3) projected axial ratio.
The position angle of the MGE is $\mathrm{PA} = 114.6$\,deg (measured from north to east).
}
\end{table}
\begin{figure}
    \centering
    \includegraphics[width=1.0\linewidth]{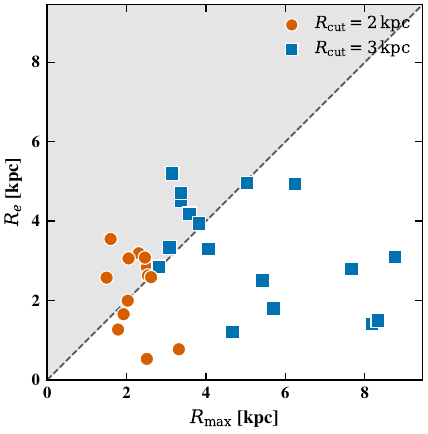}
    \caption{Galaxy effective radius \(R_e\) as a function of \(R_{\max}\), the maximum radius covered by the IFU data. The two colours indicate the radial cut adopted in the dynamical bulge definition: orange symbols correspond to \(R_{\rm cut}=2\,\mathrm{kpc}\), while blue symbols correspond to \(R_{\rm cut}=3\,\mathrm{kpc}\).
    The dashed line shows the one-to-one relation. The grey shaded region marks galaxies with \(R_e>R_{\max}\), for which the IFU coverage does not reach the galaxy effective radius.}
    \label{fig:placeholder}
\end{figure}

\begin{figure*}
\centering
\includegraphics[width=1\linewidth]{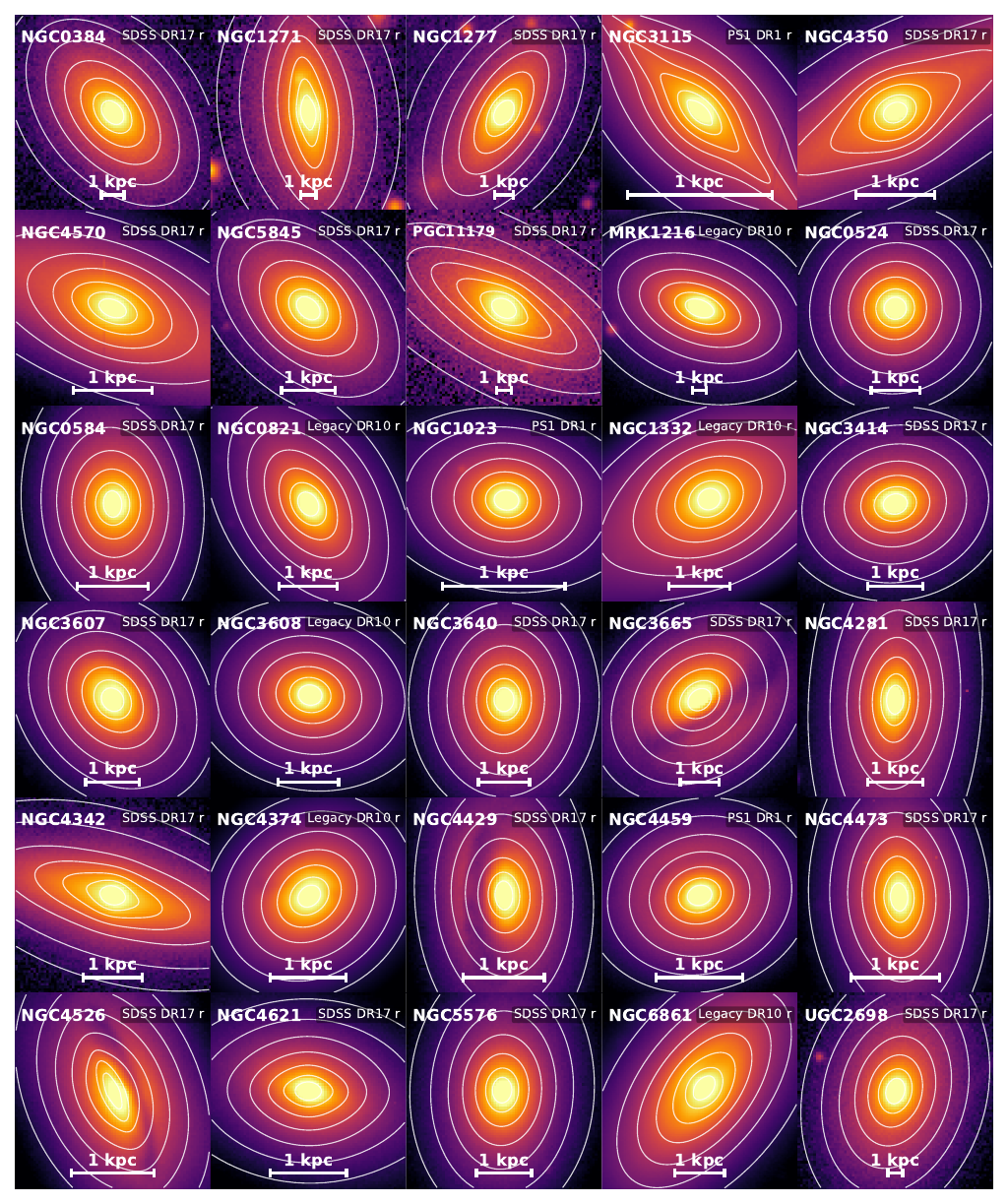}
\caption{Central $r$-band images of the 30 galaxies in our sample. Each panel covers $30\arcsec \times 30\arcsec$. The upper-right labels identify the imaging survey: Legacy denotes the DESI Legacy Imaging Surveys DR10, SDSS denotes SDSS DR17, and PS1 denotes stacked Pan-STARRS1 DR1 imaging. The white contours show the projected surface-brightness distributions reconstructed from the multi-Gaussian expansion (MGE) models adopted in our dynamical analysis. We aligned the model centres and projected major axes with the displayed images. We selected seven contour levels independently for each galaxy to sample both the central and outer structures.  The white scale bars provide a reference physical scale of 1 kpc at the adopted distance of each galaxy.}
\label{fig:25image}
\end{figure*}

\begin{figure*}[!b]
    \centering
    \includegraphics[width=16cm]{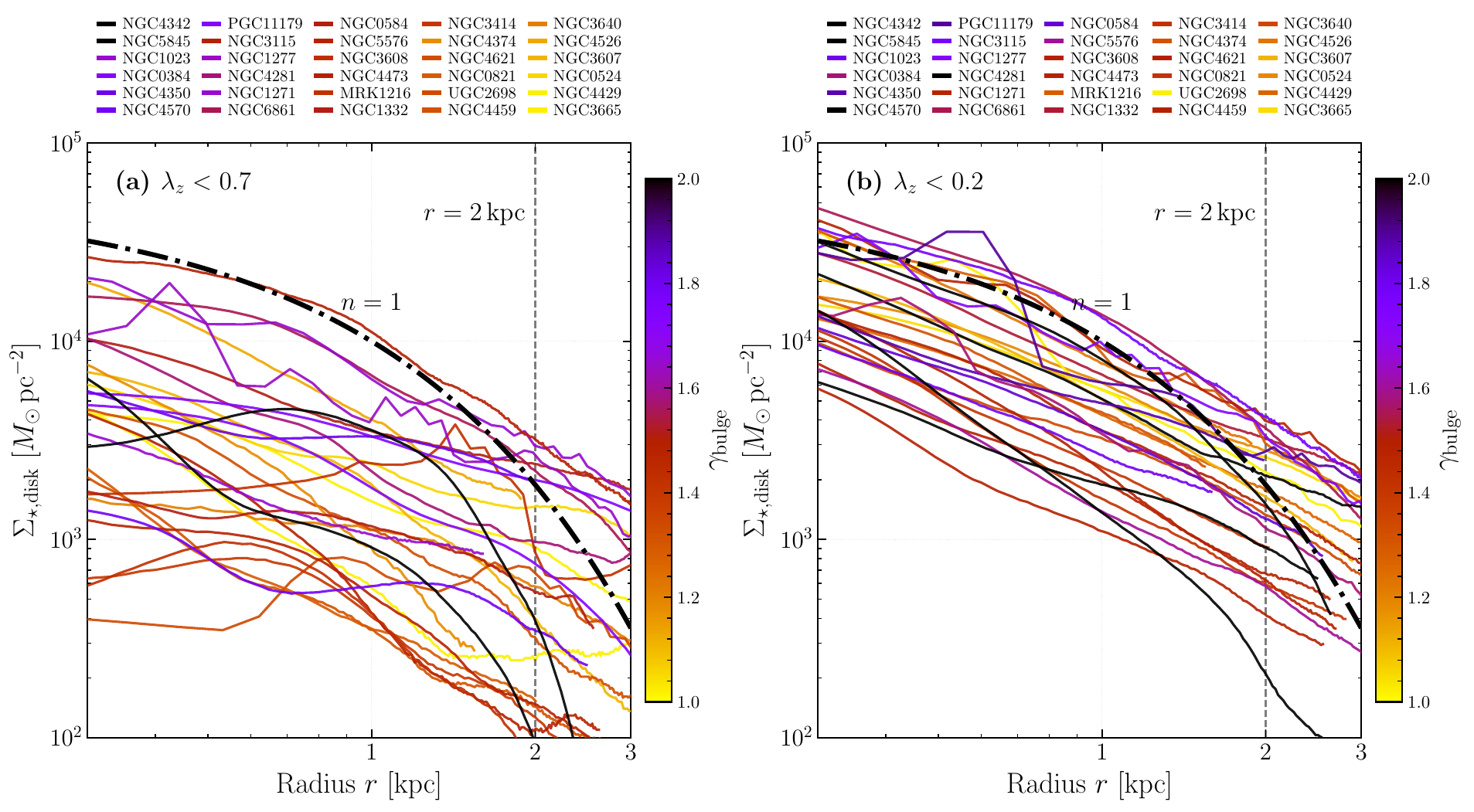}
    \caption{Same as \cref{fig:surface_density_profiles}, but for the projected disk surface mass density profiles. Panels~(a) and~(b) show the disk components associated with the two kinematic bulge definitions, $\lambda_z < 0.7$ and $\lambda_z < 0.2$, respectively. The profiles are colour-coded by the logarithmic surface-density slope of the corresponding bulge component, $\gamma_{\rm bulge}$. The dot-dashed curve shows a reference exponential S\'{e}rsic profile with $n=1$, $I_{\rm e}=10^{4}\,L_\odot\,{\rm arcsec}^{-2}$, and $R_{\rm e}=1\,{\rm kpc}$, converted to surface mass density assuming $M/L=1$.}
    \label{fig:disc_surface_density_profiles_appendix}
\end{figure*}

 \begin{figure*}
    \centering
    \includegraphics[width=12cm]{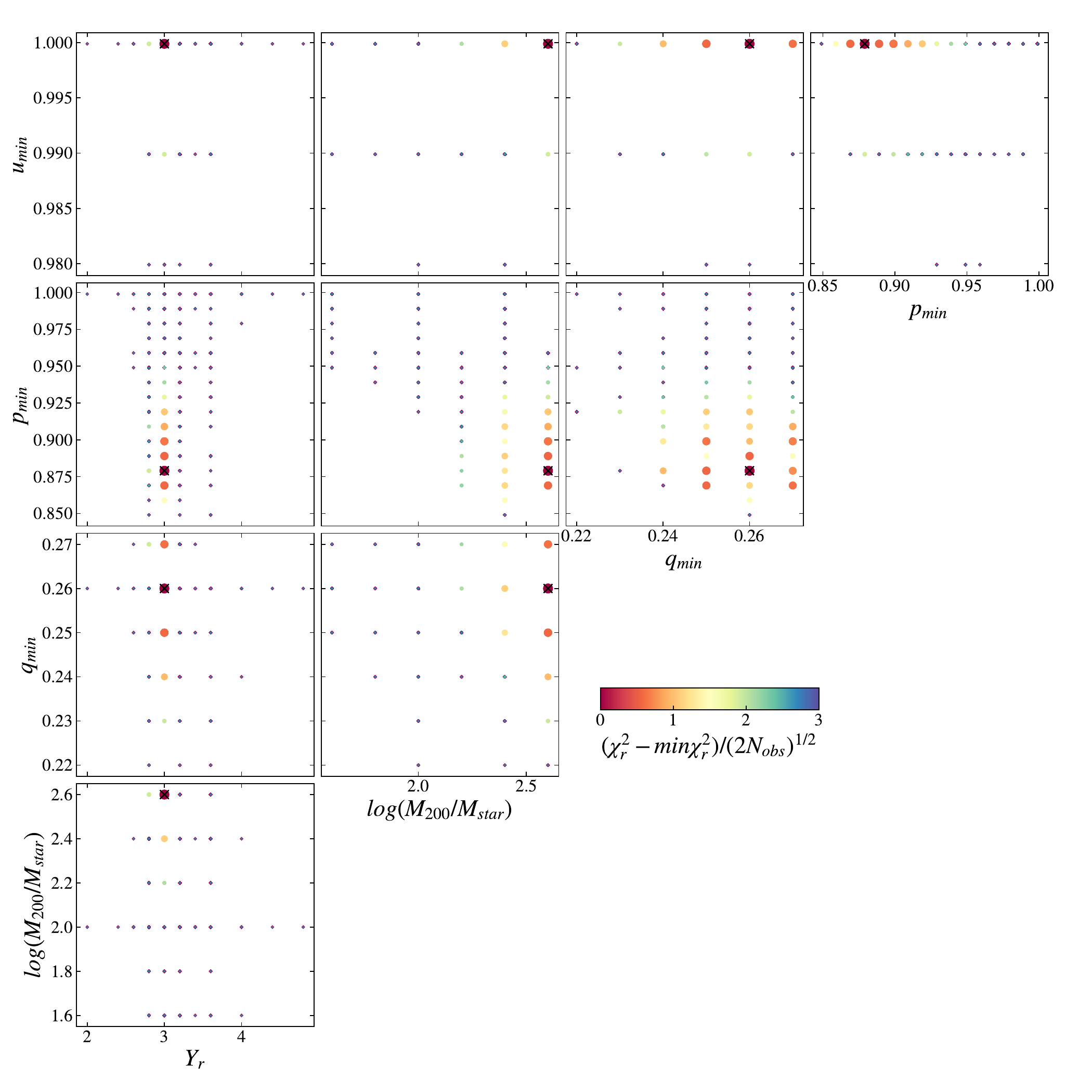}
    \caption{Exploration of the five-dimensional parameter grid for NGC~1332, shown as a representative example. The hyperparameters are: the stellar mass-to-light ratio $\Upsilon_*$ ($M_*/L$ in solar units); the dark matter halo mass $\log M_{200}$; and the intrinsic shape parameters ($p$, $q$, $u$). Each point represents a model, colour-coded according to its $\chi^2$ value as shown in the colour bar. Points satisfying $\Delta\chi^2 \equiv \chi^2 - \chi^2_{\rm min} < \sqrt{2\,N_{\rm obs}\,n_{\rm GH}}$ are within the $1\sigma$ confidence level, and are indicated by enlarged markers. The black crosses mark the best-fitting model parameters.}
    \label{fig:freepara}
\end{figure*}

\begin{figure*}
    \centering
    \includegraphics[width=16cm]{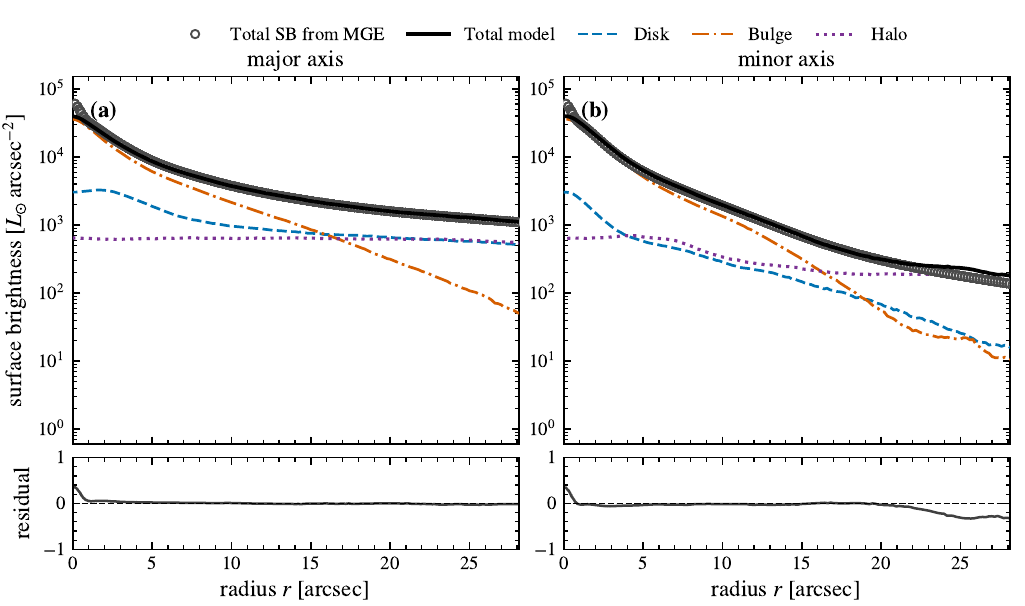}
    \caption{Surface brightness profiles of NGC~1332 along the 
    major (left) and minor (right) axes, for the $\lambda_z < 0.7$ 
    bulge definition. Open circles show the observed profiles 
    measured from the image. The black solid line shows the total 
    best-fitting model; the blue dashed, orange dash-dotted, and 
    purple dotted curves denote the disk, bulge, and halo components 
    scaled to match the observed image, respectively. Lower panels 
    show residuals between the observed profiles and the total model.}
    \label{fig:bulge_decompositon}
\end{figure*}

\end{appendix}

\end{document}